\documentclass[12pt]{article}
\usepackage{docmute}

\usepackage{epsf}
\usepackage{cite}
\usepackage{amsmath,amssymb}
\usepackage{pdfpages}
\usepackage{graphicx}
\usepackage{comment}
\usepackage{bm}
\usepackage{tikz}
\usetikzlibrary{positioning}
\usepackage{physics,cancel}
\usepackage{siunitx}
\usepackage{cleveref}

\newcommand{\mathhyphen}{\mathchar"2D}

\DeclareMathOperator\arctanh{arctanh}

\allowdisplaybreaks[1]

\crefname{section}{Section}{Section}

\newcommand{\YldYl}{Y_\ell^2}
\newcommand{\YLdYL}{Y_L^2}
\newcommand{\YRdYR}{Y_R^2}
\newcommand{\YlYld}{Y_\ell^2}
\newcommand{\YLYLd}{Y_L^2}
\newcommand{\YRYRd}{Y_R^2}
\newcommand{\lH}{\lambda_H}
\newcommand{\lHt}{\lambda_H^{\mathrm{(SM)}}}
\newcommand{\lL}{\lambda_{L}}
\newcommand{\lLt}{\lambda'_L}

\newcommand{\lR}{\lambda_R}
\newcommand{\lLR}{\lambda_{LR}}

\newcommand{\lHL}{\lambda_{HL}}
\newcommand{\lHR}{\lambda_{HR}}
\newcommand{\lkappa}{{\kappa'}}
\newcommand{\mLSq}{m_L^2}
\newcommand{\mRSq}{m_R^2}

\newcommand{\sle}[1]{\tilde{\ell}_{#1}}
\newcommand{\lam}[1]{\lambda_{#1}}
\newcommand{\lamt}[1]{\lambda'_{#1}}

\newcommand{\Seff}{S_{\rm eff}}
\newcommand{\Rsmustau}{R}

\begin{document}

\renewcommand{\thefootnote}{\fnsymbol{footnote}}
\setcounter{footnote}{0}

\begin{titlepage}

\def\thefootnote{\fnsymbol{footnote}}

\begin{center}

\hfill June, 2023\\

\vskip .5in

{\Large \bf

  Stability of Electroweak Vacuum and
  \\[1mm]
  Supersymmetric Contribution to Muon $g-2$
  \\

}

\vskip .5in

{\large
  So Chigusa$^{(a,b)}$, Takeo Moroi$^{(c)}$ and Yutaro Shoji$^{(d)}$
}

\vskip .5in

$^{(a)}$
{\em Berkeley Center for Theoretical Physics, Department of Physics,\\
University of California, Berkeley, CA 94720, USA}

\vskip 0.1in

$^{(b)}$
{\em Theoretical Physics Group, Lawrence Berkeley National Laboratory,\\
Berkeley, CA 94720, USA}

\vskip 0.1in

$^{(c)}$
{\em
Department of Physics, The University of Tokyo, Tokyo 113-0033, Japan
}

\vskip 0.1in

$^{(d)}$
{\em
Racah Institute of Physics, Hebrew University of Jerusalem, Jerusalem 91904, Israel
}

\end{center}
\vskip .5in

\begin{abstract}

  We study the stability of the electroweak vacuum in the supersymmetric
  (SUSY) standard model (SM), paying particular attention to its
  relation to the SUSY contribution to the muon anomalous magnetic
  moment $a_\mu$.  If the SUSY contribution to $a_\mu$ is sizable, the
  electroweak vacuum may become unstable because of enhanced trilinear scalar interactions in particular when the sleptons are heavy. Consequently, assuming enhanced SUSY
  contribution to $a_\mu$, an upper bound on the slepton masses is
  obtained.  We give a detailed prescription to perform a full
  one-loop calculation of the decay rate of the electroweak vacuum for the case that the SUSY
  contribution to $a_\mu$ is enhanced.  We also give an upper bound on the slepton masses as a function of the SUSY contribution to $a_\mu$.

\end{abstract}

\end{titlepage}

\renewcommand{\thepage}{\arabic{page}}
\setcounter{page}{1}
\renewcommand{\thefootnote}{\#\arabic{footnote}}
\setcounter{footnote}{0}
\renewcommand{\theequation}{\thesection.\arabic{equation}}

\section{Introduction}
\label{sec:intro}
\setcounter{equation}{0}

It has been argued that the electroweak (EW) vacuum, on which we are
living, may not be absolutely stable and it can be a false vacuum.  For example, even in the
standard model (SM) of particle physics, the EW vacuum is known to be
metastable~\cite{Isidori:2001bm, Degrassi:2012ry, Buttazzo:2013uya,
  Bednyakov:2015sca, Andreassen:2017rzq, Chigusa:2017dux,
  Chigusa:2018uuj}.  Even though the longevity of the EW vacuum is
guaranteed with the observed values of the top-quark mass $m_t \simeq
\SI{172.69}{GeV}$ and the Higgs-boson mass $m_h\simeq
\SI{125.25}{GeV}$~\cite{ParticleDataGroup:2020ssz}, the lifetime of
the EW vacuum would have been shorter than the present age of the
universe if the top-quark mass were heavier than $\sim 177-178\ {\rm
  GeV}$.  In addition, if we consider physics beyond the standard
model (BSM), the stability of the EW vacuum is not guaranteed.  This
is particularly the case in supersymmetric (SUSY) models, which have
attracted attentions as possible solutions to problems that
cannot be solved within the SM.  In particular, it has been discussed
that the SUSY contribution to the muon anomalous magnetic moment
$a_\mu$ may explain the discrepancy between the experimentally
measured value $a_\mu^{\rm (exp)}$ and the SM prediction $a_\mu^{\rm
  (SM)}$ (i.e., so-called the muon $g-2$ anomaly). For example,
adopting the SM prediction given in Ref.\ \cite{Aoyama:2020ynm}, the
deviation between $a_\mu^{\rm (exp)}$ and $a_\mu^{\rm (SM)}$ is about
$4.2\sigma$ (for more detail, see the next section). It has been
discussed that the SUSY contribution can be large enough to explain
such a deviation (see, for example, \cite{Chakraborti:2021kkr,
  Endo:2021zal, Han:2021ify, VanBeekveld:2021tgn, Ahmed:2021htr,
  Cox:2021nbo, Wang:2021bcx, Baum:2021qzx, Yin:2021mls,
  Iwamoto:2021aaf, Athron:2021iuf, Shafi:2021jcg, Aboubrahim:2021xfi,
  Chakraborti:2021bmv, Baer:2021aax, Aboubrahim:2021phn, Li:2021pnt,
  Jeong:2021qey, Ellis:2021zmg, Nakai:2021mha, Forster:2021vyz,
  Ellis:2021vpp, Chakraborti:2021mbr, Gomez:2022qrb,
  Chakraborti:2022vds, Agashe:2022uih, Morrison:2022vqe, Li:2022zap,
  Zhao:2022pnv, He:2023lgi} for recent studies). 

In the minimal SUSY SM (MSSM), there is an extra SUSY contribution to the muon anomalous
magnetic moment, which can mitigate the muon $g-2$ anomaly.  Because the
SUSY contribution to the muon $g-2$ is due to the diagrams with
superparticles in the loops, it is suppressed as the superparticles
become heavier.  Thus, in order to explain the muon $g-2$ anomaly,
the masses of superparticles are bounded from above.  A detailed
understanding of the upper bound is important in order to test the
SUSY interpretation of the muon $g-2$ anomaly with the on-going and the future
collider experiments \cite{Endo:2013lva, Endo:2013xka, Endo:2022qnm,
  Chigusa:2022xpq}.  The muon $g-2$ anomaly can be explained in
various parameter regions of the MSSM.  If the masses of all the
superparticles are comparable, they are
required to be of $O(100)\ {\rm GeV}$.  Then, the muon $g-2$ anomaly
indicates that superparticles (in particular, sleptons, charginos, and
neutralinos) are important targets of the on-going and the future collider
experiments.  The SUSY contribution to the muon $g-2$ can be, however,
sizable even if some of the superparticles are much heavier.  It happens when the
Higgsino mass parameter (i.e., the so-called $\mu$ parameter) is
significantly large, which results in the enhanced smuon-smuon-Higgs trilinear scalar
coupling.

The enhanced scalar trilinear couplings may cause the EW vacuum
instability.  In the MSSM, scalar partners of SM fermions (i.e.,
quarks and leptons) are introduced.  As we will discuss in the
following sections, the absolute minimum of the scalar potential in
the MSSM may be the color- and/or charge-breaking (CCB) minimum at which
sfermions acquire non-zero expectation values.  When the potential has
the CCB absolute minimum, the EW vacuum is a false vacuum and decays
into the true vacuum (i.e., CCB one) with a finite lifetime; the model is not viable if the
lifetime is shorter than the present age of the Universe.  In
particular, as the scalar trilinear couplings become larger, the
corresponding CCB vacuum tends to have lower potential energy and the
EW vacuum becomes less stable \cite{Frere:1983ag, Gunion:1987qv,
  Casas:1995pd, Kusenko:1996jn}.  Thus, requiring that $a_{\mu}^{\rm
  (SUSY)}$ be large enough to solve (or relax) the muon $g-2$ anomaly,
we can obtain upper bounds on the masses of superparticles based on
the observed longevity of the EW vacuum.  

The stability of the EW vacuum was considered in Ref.\ \cite{Endo:2021zal} in connection with the muon $g-2$ anomaly. Ref.\ \cite{Endo:2021zal} estimated the lifetime of the EW vacuum using the tree-level bounce action. (For other studies of the stability of the EW vacuum in
the MSSM, see Refs.\ \cite{Endo:2013lva,Chowdhury:2013dka,Badziak:2014kea,Duan:2018cgb,Hollik:2018wrr,Chigusa:2022xpq}.  The tree-level estimation
of the decay rate, however, suffers from the uncertainty in
determining the normalization factor of the decay rate (i.e., the
prefactor $\mathcal{A}$ which will be introduced in Section
\ref{sec:vacuumdecay}).  In order for a reliable estimate of the
decay rate, the ab-initio calculation of the prefactor is necessary,
which requires a full one-loop calculation of the decay rate.  Results of such a calculation were presented in
Ref.\ \cite{Chigusa:2022xpq} by the present authors, leaving the
detailed formulas for the calculation to the subsequent publication.
In this paper, in the following, we give a detailed prescription to
perform such a one-loop calculation, based on the state-of-the-art
formula to calculate the decay rate of a false vacuum
\cite{Endo:2017gal, Endo:2017tsz, Chigusa:2020jbn}.

The aim of this paper is to provide a detailed description of the one-loop calculation of the vacuum decay rate in the MSSM.  In this paper, we present explicit formulas based on which the one-loop calculation can be performed. Then, we extend the analysis of Ref.\ \cite{Chigusa:2022xpq} and derive an upper bound on the smuon mass as a function of the SUSY contribution to the muon anomalous magnetic moment.
We also study the case where three flavors of the sleptons and the Bino are relatively light, while all the other superparticles are decoupled.

This paper is organized as follows.  In \cref{sec:mssm}, we overview
the SUSY contribution to the muon anomalous magnetic moment.  In
\cref{sec:eft}, we define low-energy effective field theories (EFTs),
which are obtained by integrating out heavy superparticles irrelevant
to our discussion.  The EFTs introduced in \cref{sec:eft} are used
for the calculation of the decay rate.  In
\cref{sec:vacuumdecay}, we present the prescription to perform the
one-loop calculation of the decay rate, taking into account the
effects of SUSY particles relevant for the muon $g-2$ anomaly. In
\cref{sec:results}, we discuss phenomenological implications of the EW
vacuum stability by numerically evaluating the decay rate of the EW
vacuum.  \cref{sec:conclusions} is devoted to the conclusion and
discussions.

\section{Muon $g-2$: SM and SUSY Contributions}
\label{sec:mssm}
\setcounter{equation}{0}

In this section, we first summarize the SM prediction of $a_\mu$
as well as the experimentally measured value.  Then, we give a brief
overview of the SUSY contributions.

\subsection{SM prediction}

The muon anomalous magnetic moment has been measured with very high
accuracy.  Combining the results of BNL and FermiLab experiments, the
experimentally measured value of the muon anomalous magnetic moment
$a_\mu$ is given by \cite{Bennett:2002jb, Bennett:2004pv, Bennett:2006fi, Abi:2021gix}
\begin{align}
  a_\mu^{\rm (exp)} =
  (11\,659\,206.1 \pm 4.1 ) \times 10^{-10}.
  \label{amu(exp)}
\end{align}

Significant efforts have been made to understand the theoretical
prediction of the muon anomalous magnetic moment.  In particular, in
the SM, a very precise calculation of $a_\mu$ has been performed.  One important quantity necessary to
obtain the theoretical prediction is the hadronic vacuum
polarization (HVP) of photon.  The effect of the HVP has been
estimated by using the so-called $R$-ratio from the data provided by
$e^+e^-$ collider experiments; combining the HVP contribution based on
the $R$-ratio with other contributions,
Ref.\ \cite{Aoyama:2020ynm} obtained the SM prediction as
\begin{align}
  a_ \mu^{\rm (SM)} = ( 11\,659\,181.0 \pm 4.3 ) \times 10^{-10}.
  \label{amu(SM)}
\end{align}
We adopt the above result as our canonical value of the SM prediction.

Comparing Eqs.\ \eqref{amu(exp)} and \eqref{amu(SM)}, the
experimentally measured value is about $4.2\sigma$ away from the SM
prediction:
\begin{align}
  a_\mu^{\rm (exp)} - a_\mu^{\rm (SM)} = ( 25.1 \pm 5.9) \times 10^{-10}.
  \label{damu}
\end{align}
The deviation is sometimes called the muon $g-2$ anomaly.  For
later convenience, we define the ``$0\sigma$,'' ``$1\sigma$,'' and
``$2\sigma$'' values of BSM contribution to $a_\mu$ necessary to
resolve the discrepancy:
\begin{align}
  \Delta a_{\mu}^{(0\sigma)} = 25.1 \times 10^{-10},
  \\
  \Delta a_{\mu}^{(1\sigma)} = 19.2 \times 10^{-10},
  \\
  \Delta a_{\mu}^{(2\sigma)} = 13.3 \times 10^{-10}.
\end{align}

Besides, the HVP has been also estimated by using lattice Monte Carlo
simulation.  The BMW collaboration reached the sub-percent precision
in calculating the leading-order HVP to $a_\mu$
\cite{Borsanyi:2020mff}, based on which the tension between the
experimentally measured value and the SM prediction is significantly
weakened.  Other recent results of lattice calculations are consistent
with the BMW result, particularly for the so-called ``intermediate
time window observable'' \cite{Wang:2022lkq, Ce:2022kxy,
  ExtendedTwistedMass:2022jpw, Bazavov:2023has, Blum:2023qou}.
Adopting the BMW result for the estimation of the HVP contribution,
$a_\mu^{\rm (exp)}-a_\mu^{\rm (SM, BMW)}=(10.7\pm 7.0)\times 10^{-10}$
(with $a_\mu^{\rm (SM, BMW)}$ being the SM prediction based on the BMW
result), meaning that the consistency between $a_\mu^{\rm (exp)}$ and
$a_\mu^{\rm (SM, BMW)}$ is $1.5\sigma$ \cite{latticeave2023}. In
addition,
a new measurement of the cross section of the process
$e^+e^-\rightarrow\pi^+\pi^-$ has been performed in the center of mass
energy range from $0.32$ to $1.2\ {\rm GeV}$
using the CMD-3 detector 
at the $e^+e^-$ collider VEPP-2000 \cite{CMD-3:2023alj}; its
result indicates $\sim 3\ \%$ increase of the $\pi^{+} \pi^{-}$ contribution to the HVP contribution to $a_\mu$
compared to the one before the CMD-3 result.
If one uses the value given in Ref.\ \cite{CMD-3:2023alj}, the tension
between the theoretical prediction and the experimentally measured value
of $a_\mu$ reduces to $\sim 2.3\ \sigma$.  Currently, it is
still premature to conclude which value of the SM prediction of
$a_\mu$ is the most reliable.  In this paper, we take
Eq.\ \eqref{damu} seriously and regard the discrepancy as a hint of
the BSM physics, although we will also comment on the implications of
the lattice results of the HVP contribution on the stability of the EW
vacuum in the MSSM.

\subsection{Supersymmetric contribution to muon $g-2$}

As discussed in the previous subsection, the experimentally measured
value of the muon anomalous magnetic moment may be in tension with the
SM prediction. The discrepancy may suggest the existence of BSM physics
that affects $a_\mu$.  SUSY
is one of such possibilities and, in the following, we assume that it
is the case.  Below, we overview the SUSY contribution to
the muon $g-2$ in the MSSM. We also explain why the
stability of the EW vacuum is important in the study of the SUSY
contribution.

It is well known that the SUSY contribution to the muon anomalous
magnetic moment, which is denoted as $a_{\mu}^{\rm (SUSY)}$, is
enhanced when $\tan\beta$ is large. Here $\tan\beta$ is the ratio of the two Higgs bosons
(denoted as $H_u$ and $H_d$, respectively).  We concentrate on the
large $\tan\beta$ case so that the requirement on the mass scale of
the smuons to solve the muon $g-2$ anomaly is expected to become higher.
Notice that $\tan\beta$ cannot be arbitrarily large if we require the
perturbativity of the coupling constants.  In particular, because the
grand unified theory (GUT) is one of the strong motivations to
consider the MSSM, we require that the coupling constants (in
particular, the bottom Yukawa coupling constant) are perturbative up
to the GUT scale.  Then, $\tan\beta$ should be smaller than
$\sim 50-60$.

In order to study the behavior of $a_{\mu}^{\rm (SUSY)}$ in the large
$\tan\beta$ case, it is instructive to use the so-called mass
insertion approximation in which $a_{\mu}^{\rm (SUSY)}$ is estimated
in the gauge-eigenstate basis with treating the interactions
proportional to the VEVs of the Higgs bosons as perturbations.  (In our
following numerical calculation, however, $a_{\mu}^{\rm (SUSY)}$ is estimated
more precisely using the mass eigenstates of the sleptons, charginos,
and neutralinos, as we will explain.)
Because the superparticles are in the loop, $a_{\mu}^{\rm (SUSY)}$ is
suppressed as the superparticles become heavier.  For the case where
the masses of all the superparticles are comparable, for example, the
SUSY contribution to the muon anomalous magnetic moment is
approximately given by $|a_\mu^{\rm (SUSY)}|\simeq
\frac{5g_2^2}{192\pi^2} \frac{m_\mu^2}{m_{\rm SUSY}^2}\tan\beta$,
where $g_2$ is the gauge coupling constant of $SU(2)_L$ and $m_{\rm
  SUSY}^2$ is the mass scale of superparticles.  (Here, the
contributions of the diagrams containing the Bino are neglected
because they are subdominant.)  Taking $\tan\beta\sim 50$, the
superparticles should be lighter than $\sim 700\ {\rm GeV}$ in order
to make the total muon anomalous magnetic moment consistent with the
observed value at the $2\sigma$ level.

\begin{figure}[t]
  \centering \includegraphics[height=0.15\textheight]{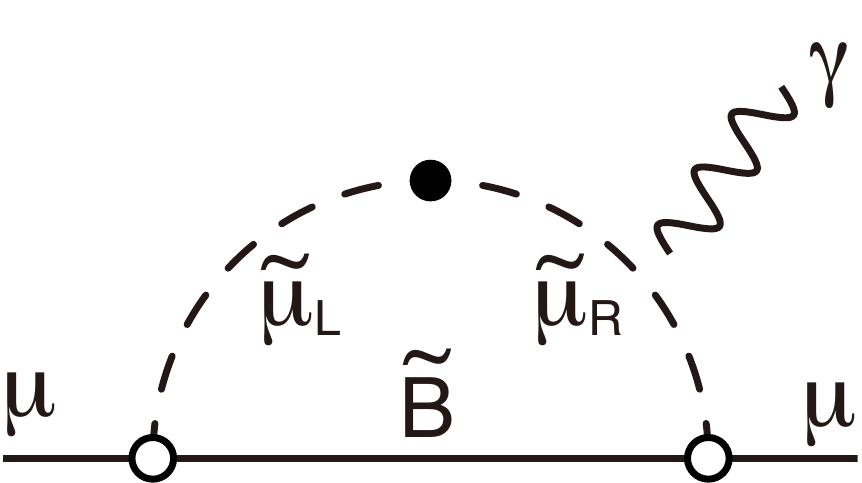}
  \caption{Bino-Smuon diagram of the SUSY contribution to the muon
  anomalous magnetic moment.  The mass insertion approximation is
  adopted.  The black blob is the two-point interaction induced by the
  VEVs of the Higgs bosons.}  \label{fig:feyndiag}
\end{figure}

Such an upper bound is significantly altered by the Bino-smuon diagram
(see Fig.\ \ref{fig:feyndiag}).  The other $\tan\beta$-enhanced
diagrams have slepton, gaugino, and Higgsino propagators in the loop
and hence contributions of them are suppressed when any of the
sleptons, gauginos, or Higgsinos are heavy.  On the contrary, the
Bino-smuon diagram has only the smuon and Bino propagators in the
loop, and its contribution is approximately proportional to the
Higgsino mass parameter, $\mu$.  Thus, in the parameter region with a
very large $\mu$ parameter, the contribution of the Bino-smuon diagram
can be large enough to cure the muon $g-2$ anomaly even if some of the
superparticles (in particular, Higgsinos) are much heavier than the
upper bound estimated above.

In the following, we study the upper bound on the masses of
superparticles in the light of the muon $g-2$ anomaly, paying
particular attention to the contribution of the Bino-smuon diagram.
In a parameter region where the
Bino-smuon diagram dominates $a_\mu^{\rm (SUSY)}$, the $\mu$ parameter is large and hence the
smuon-smuon-Higgs trilinear coupling is enhanced.  A large trilinear
scalar coupling is, in general, dangerous because it may make the EW vacuum unstable.  In the present case, a large smuon-smuon-Higgs
trilinear coupling develops another deeper vacuum, with which the EW vacuum becomes a false vacuum.
Consequently, the lifetime of the EW vacuum may become shorter than the present cosmic
age \cite{ParticleDataGroup:2020ssz}:
\begin{align}
  t_{\rm now} \simeq 13.8\ {\rm Gyr}.
  \label{t_now}
\end{align}
In addition, a large $\mu$ parameter generally enhances all the
slepton-slepton-Higgs trilinear couplings. When the stau is light, it is important to study the
instability due to the stau-stau-Higgs trilinear coupling since it typically gives a more
stringent constraint than the smuon-smuon-Higgs trilinear coupling.

\section{Effective Field Theory Analysis}
\label{sec:eft}
\setcounter{equation}{0}

\subsection{MSSM}

Our main purpose is to calculate the decay rate of the EW vacuum,
taking into account the effects of SUSY particles.  In particular, we
are interested in the upper bound on the masses of superparticles
under the requirement that the muon $g-2$ anomaly be solved (or
relaxed) by the SUSY contribution.  We are interested in the case where the $\mu$ parameter is
large so that the Bino-smuon diagram dominates $a_\mu^{\rm (SUSY)}$;
hereafter, we focus on the case of $\mu\gtrsim O(10)\ {\rm TeV}$.  In
such a case, $a_\mu^{\rm (SUSY)}$ can be large enough to solve the
muon $g-2$ anomaly even with relatively large smuon and Bino masses.

A large value of $\mu$ implies heavy Higgsinos.
In addition, in order to push the lightest Higgs mass up to the observed value, i.e., about $125\ {\rm GeV}$ \cite{ATLAS:2015yey, ATLAS:2018tdk, CMS:2020xrn}, relatively heavy stop masses are preferred \cite{Bagnaschi:2014rsa}, which are characterized by a scale $M_S$.
On the contrary, in order to enhance $a_\mu^{\rm (SUSY)}$, smuon and Bino masses should be close to the EW scale.
Based on these considerations, in this paper, we consider two cases where the Bino, denoted by $\tilde{B}$, and either only the second generation of sleptons or all three generations of sleptons are relatively light among the MSSM particles.
The other MSSM constituents are assumed to have masses of $O(M_S)$.
Here, we assume that Wino and gluino masses are of $O(M_S)$ just for simplicity.
We also assume that there exists a large hierarchy between the EW scale and $M_S$.

A large value of $\mu$ suggests relatively large values of the soft
SUSY breaking Higgs mass parameters for a viable EW symmetry
breaking; the lighter Higgs mass, i.e., the mass of the SM-like Higgs
boson $H$, is realized by the cancellation between the contributions of
the $\mu$ and soft SUSY breaking parameters.  The heavier Higgs bosons are expected to have masses of $O(M_S)$, comparable to $\mu$.
In such a case, the SM-like Higgs $H$ and the heavy doublet
$H'$ are given by linear combinations of the up- and down-type
Higgs bosons as
\begin{align}
  \left( \begin{array}{c} H \\ H' \end{array} \right)
  = 
  \left( \begin{array}{cc} 
    \cos\beta & \sin\beta
    \\
    -\sin\beta & \cos\beta
  \end{array} \right)
  \left( \begin{array}{c} H_d^*\\ H_u \end{array} \right).
\end{align}

In summary, we are interested in the case where the mass spectrum below $\sim$ TeV scale 
includes one or three sleptons and the Bino $\tilde{B}$, as well as the SM
particles.  Hereafter, the leptons (sleptons) in the $\alpha$-th generation in the
gauge eigenstate are denoted as $\ell_{L\alpha}$ and $\ell_{R\alpha}$ ($\tilde{\ell}_{L\alpha}$ and $\tilde{\ell}_{R\alpha}$); $\ell_{L\alpha}$ and $\tilde{\ell}_{L\alpha}$ are $SU(2)_L$ doublets with hypercharge $-\frac{1}{2}$, while $\ell_{R\alpha}$ and $\tilde{\ell}_{R\alpha}$ are $SU(2)_L$ singlets with hypercharge $+1$.
For the calculation of the muon $g-2$, sleptons in the second generation are important; they are also denoted as
\begin{align}
  \tilde{\ell}_{L2} =
  \begin{pmatrix}
    \tilde{\nu}_\mu \\ \tilde{\mu}_L
  \end{pmatrix},~~~
  \tilde{\ell}_{R2} = \tilde{\mu}_R.
\end{align}

\subsection{EFT}

We consider the case where there exists a hierarchy between the masses of MSSM particles.  To properly evaluate the coupling constants in such a case, we resort to the EFT approach; we use the renormalization group (RG) analysis to evaluate the EFT parameters at the EW scale.

We adopt the top-quark pole mass $M_t$ and $M_S$ as matching scales of
different EFTs.  For the renormalization scale $Q < M_t$, we consider
the QCD+QED that contains the SM gauge couplings and fermion masses as
parameters.  For $M_t < Q < M_S$, we adopt the EFT containing the Bino and the
sleptons in addition to all the SM particles, which we call the slepton
EFT (see below).  At $Q=M_S$, the slepton EFT is matched to the MSSM.
This matching is necessary to relate scalar couplings in the
slepton EFT to gauge and Yukawa couplings.  We define $M_S$ through
the following procedure.  At $Q=M_t$, we give a tree-level estimation of
couplings through the relations described below, with which we
estimate the Higgsino mass so that the required size of the muon $g-2$ can be obtained from the MSSM contribution. We use this estimate of the Higgsino mass as $M_S$.

To specify the model parameters relevant to the calculation of the
decay rate of the EW vacuum, we show the Lagrangian of the slepton EFT, which is
obtained by integrating out the heavy MSSM particles.  In the following,
for simplicity, we assume that the effects of CP and flavor violations
are negligible.  Then, the EFT Lagrangian is given by
\begin{align}
  \mathcal{L} = \mathcal{L}_{\mathrm{kin}}^{(\mathrm{SM})}
  + \Delta \mathcal{L}_{\mathrm{kin}}
  + \Delta \mathcal{L}_{\mathrm{Yukawa}} - V,
  \label{eq:EFT_Lagrangian}
\end{align}
where $\mathcal{L}_{\mathrm{kin}}^{(\mathrm{SM})}$ is the kinetic
terms of the SM fields (including the gauge interactions), while
$\mathcal{L}_{\mathrm{Yukawa}}^{(\mathrm{SM})}$ represents the SM-like
Yukawa interactions.  The latter includes the lepton Yukawa
interactions as
\begin{align}
  \mathcal{L}_{\mathrm{Yukawa}}^{\mathrm{(SM)}} \ni 
  Y_{\ell \alpha} \ell_{L\alpha}^\dagger H \ell_{R\alpha}
  + \mathrm{h.c.},
\end{align}
where $\alpha$ is the flavor index and the sum over the flavor indices
is implicit.  The additional kinetic terms and Yukawa couplings are
described by
\begin{align}
  \Delta \mathcal{L}_{\mathrm{kin}} &=
  | D_\mu \sle{L\alpha} |^2 + | D_\mu \sle{R\alpha} |^2
  - i \tilde{B} \sigma^\mu \partial_\mu \tilde{B}^\dagger
  - \left(
    \frac{1}{2} M_1 \tilde{B}\tilde{B} + \mathrm{h.c.}
  \right), \\
  \Delta \mathcal{L}_{\mathrm{Yukawa}} &= 
  Y_{R\alpha} \sle{R\alpha}^\dagger \ell_{R\alpha} \tilde{B}^\dagger
  + Y_{L\alpha} \sle{L\alpha}^\dagger \ell_{L\alpha} \tilde{B}
  + \mathrm{h.c.},
\end{align}
where $D_\mu$ denotes the covariant derivative.  The scalar potential $V$ is given by
\begin{align}
  V = &\, V_2 + V_3 + V_4,
\end{align}
where
\begin{align}
  V_2 = &\, m_H^2 |H|^2
  + m_{R\alpha}^2\, |\sle{R\alpha}|^2
  + m_{L\alpha}^2\, |\sle{L\alpha}|^2,
  \label{eq:V2} \\
  V_3 = &\, - T_{\alpha} \sle{R\alpha}^{*} H^\dagger \sle{L\alpha}
  + \text{h.c.},
  \\
  V_4 = &\, \lambda_H |H|^4
  + \lam{HR\alpha} |H|^2 |\sle{R\alpha}|^2
  + \lam{HL\alpha} |H|^2 |\sle{L\alpha}|^2
  + \kappa_{\alpha} 
  ( H^\dagger \sle{L\alpha} ) 
  ( \sle{L\alpha}^\dagger H )
  \notag \\ &
  + \lam{R\alpha\beta} | \sle{R\alpha} |^2 | \sle{R\beta} |^2
  + \lam{L\alpha\beta} | \sle{L\alpha} |^2 | \sle{L\beta} |^2
  + \lamt{L\alpha\beta} | \sle{L\alpha}^\dagger \sle{L\beta} |^2
  + \lam{LR\alpha\beta} | \sle{R\alpha} |^2 | \sle{L\beta} |^2. 
  \label{eq:V4}
\end{align}
Note that there are some redundancies in the choice of the coupling constants.
To resolve these redundancies, we work under the convention that $\lam{R\alpha\beta}=\lam{R\beta\alpha}$, $\lam{L\alpha\beta}=\lam{L\beta\alpha}$, $\lamt{L\alpha\beta}=\lamt{L\beta\alpha}$, and $\lamt{L\alpha\alpha}=0$. 
The first flavor index of $\lam{LR\alpha\beta}$ is for the right-handed sleptons and the second one is for the left-handed sleptons.
For the sake of the following discussion, we also define
\begin{align}
  \kappa'_{\alpha} \equiv \kappa_\alpha + \lam{HL\alpha}.
\end{align}

Coupling constants in different EFTs are related to each other through
the matching conditions at the threshold scales.  All
the SM parameters including the Higgs quartic coupling $\lHt$ and the mass
$m_H^{2\mathrm{(SM)}}$ are determined at the energy scale $M_t$ and
below.  Some of the couplings such as the top Yukawa
coupling, gauge couplings, $\lHt$, and $m_H^{2\mathrm{(SM)}}$ are
significantly affected by the weak-scale threshold corrections.  We
use the results of Ref.\ \cite{Buttazzo:2013uya} to fix these
parameters with using physical parameters $\alpha_3(M_Z)=0.1179$,
$M_t=172.76\,\mathrm{GeV}$, $M_W=80.379\,\mathrm{GeV}$, and
$M_h=125.25\,\mathrm{GeV}$ \cite{ParticleDataGroup:2020ssz} as inputs.
As for the light fermion couplings, we calculate the running of their
masses with the one-loop QED and the three-loop QCD beta functions
\cite{Gorishnii:1990zu, Tarasov:1980au, Gorishnii:1983zi} to determine
the corresponding Yukawa couplings at $Q=M_t$.

We perform the matching between the SM and the slepton EFT at $Q=M_t$ taking into account important one-loop corrections.  For the Higgs
quartic coupling and the mass term, we adopt
\begin{align}
  \lHt &= \lH - \Delta \lH, \label{eq:dellH} \\
  m_H^{2\mathrm{(SM)}} &= m_H^2 - \Delta m_H^2, \label{eq:delmHSq}
\end{align}
with
\begin{align}
  (16\pi)^2 \Delta \lH =& \left(
  \lam{HL\alpha}^2 + \frac{1}{2} {\kappa'_\alpha}^2 + \lam{HL\alpha} \kappa'_\alpha
  \right) B_0(m_{L\alpha}^2, m_{L\alpha}^2)
  + \frac{1}{2} \lam{HR\alpha}^2 B_0(m_{R\alpha}^2, m_{R\alpha}^2) \notag 
  \\ &
  + \lam{HL\alpha} T_{\alpha}^2 C_0 (m_{L\alpha}^2, m_{L\alpha}^2, m_{R\alpha}^2)
  + \lam{HR\alpha} T_{\alpha}^2 C_0 (m_{R\alpha}^2, m_{R\alpha}^2, m_{L\alpha}^2)
  \notag \\ &
  + \frac{1}{2} T_{\alpha}^4 D_0(m_{L\alpha}^2, m_{R\alpha}^2, m_{L\alpha}^2, m_{R\alpha}^2),
  \\
  (16\pi)^2 \Delta m_H^2 =&
  \left( 2\lam{HL\alpha} + \kappa'_\alpha \right) A_0(m_{L\alpha}^2)
  + \lam{HR\alpha} A_0(m_{R\alpha}^2)
  + T_{\alpha}^2 B_0 (m_{L\alpha}^2, m_{R\alpha}^2),
\end{align}
where $A_0$, $B_0$, $C_0$, and $D_0$ are the Passarino-Veltman one-,
two-, three-, and four-point functions without momentum inflow
\cite{Passarino:1978jh}, respectively.  We also take account of the
one-loop corrections to the lepton Yukawa couplings because it can
significantly affect the decay rate of the EW vacuum.  The one-loop correction
to the lepton Yukawa matrix $\Delta Y_\ell$, with which the EFT Yukawa
matrix $Y_\ell$ is related to the SM one $Y_\ell^{\mathrm{(SM)}}$ as 
$Y_\ell^{\mathrm{(SM)}}=Y_\ell-\Delta Y_\ell$, is given by
\begin{align}
  (16\pi)^2 \Delta Y_{\ell\alpha} = Y_{R\alpha} T_{\alpha} Y_{L\alpha} M_1 J(M_1^2, m_{R\alpha}^2, m_{L\alpha}^2),
\end{align}
with
\begin{align}
  J(a,b,c) \equiv -\frac
  {ab\ln(a/b) + bc\ln(b/c) + ca\ln(c/a)}
  {(a-b)(b-c)(c-a)}. \label{eq:I}
\end{align}
These corrections can be sizable because of the hierarchy between $M_t$ and $M_S$.

Some of the parameters in the slepton EFT are related to each other due to the SUSY relation among coupling constants. Thus, we should impose the matching conditions on the couplings at the matching scale $Q=M_S$.
At the tree-level, these conditions are given by
\begin{align}
  Y_{L\alpha} &= \sqrt{\frac{3}{10}} g_1,\\
  Y_{R\alpha} &= -\sqrt{\frac{6}{5}} g_1,
\end{align}
for the slepton Yukawa couplings, and
\begin{align}
  \lam{R\alpha\beta} &= \frac{3}{10} g_1^2,\\
  \lam{L\alpha\beta} &= \frac{1}{8} g_2^2 (2\delta_{\alpha\beta}-1) + \frac{3}{40} g_1^2,\\
  \lamt{L\alpha\beta} &= \frac{1}{4} g_2^2 (1-\delta_{\alpha\beta})\\
  \lam{LR\alpha\beta} &= \frac{|Y_{\ell\alpha}|^2}{\cos^2 \beta} \delta_{\alpha\beta} - \frac{3}{10} g_1^2, \label{eq:lLR} \\
  \lam{HR\alpha} &= |Y_{\ell\alpha}|^2 - \frac{3}{10} g_1^2 \cos 2\beta, \label{eq:lHRm} \\
  \lam{HL\alpha} &= \left( \frac{1}{4} g_2^2 + \frac{3}{20} g_1^2 \right) \cos 2\beta,\\
  \kappa'_\alpha &= |Y_{\ell\alpha}|^2 - \left( \frac{1}{4} g_2^2 - \frac{3}{20} g_1^2 \right) \cos 2\beta, \label{eq:lkappam} \\
  T_\alpha &= Y_{\ell\alpha} \mu \tan\beta, \label{eq:TBC}
\end{align}
for the scalar quartic and trilinear couplings, where $g_2$ and $g_1$ are
the gauge couplings of $SU(2)_L$ and $U(1)_Y$, respectively.  (Here we
use the $SU(5)$ normalization of the $U(1)_Y$ coupling $g_1^2 \equiv
5g_Y^2 / 3$.)
We neglect SUSY breaking trilinear scalar interactions for simplicity.  We also neglect the threshold corrections to the above matching conditions due to the MSSM particles as heavy as $\sim M_S$, because such corrections depend on the detailed mass spectrum of heavy particles.

Although we determine $\lH$ at $Q=M_t$, there is also a SUSY relation between $\lH$ and other couplings.
Considering only the stop contribution to the threshold correction, which is in many cases the largest, we obtain the one-loop matching condition \cite{Bagnaschi:2014rsa}
\begin{align}
  \lambda_H = \left[ \frac{1}{8} g_2^2 + \frac{3}{40} g_1^2 \right] \cos^2 2\beta
  + \delta \lambda_H,
  \label{eq:lambdaH_matching}
\end{align}
with
\begin{align}
  (16\pi^2) \delta \lambda_H \simeq &\, \frac{3}{2} y_t^2 \left[ y_t^2 + \left( \frac{1}{2} g_2^2 - \frac{1}{10} g_1^2 \right) \cos2 \beta \right]
  \ln \frac{m_{Q3}^2}{Q^2} \notag \\
  &+ \frac{3}{2} y_t^2 \left[ y_t^2 + \frac{2}{5} g_1^2 \cos2 \beta \right]
  \ln \frac{m_{U3}^2}{Q^2} \notag \\
  &+ \frac{\cos^2 2\beta}{200} \left[ (25g_2^4 + g_1^4) \ln \frac{m_{Q3}^2}{Q^2}
  + 8 g_1^4 \ln \frac{m_{U3}^2}{Q^2}
  + 2 g_1^4 \ln \frac{m_{D3}^2}{Q^2} \right],
\end{align}
where $y_t$ is the top Yukawa coupling, and $m_{Q3}$, $m_{U3}$, and $m_{D3}$ the mass parameters of the third generation left-handed squark, right-handed up-type squark, and right-handed down-type squark, respectively.
Once the value of $\lH$ at the matching scale $Q=M_S$ is obtained, we can solve \eqref{eq:lambdaH_matching} against the stop mass $m_{\tilde{t}}$ assuming the universality $m_{\tilde{t}} \equiv m_{Q3} = m_{U3} = m_{D3}$.
It is known that a sizable threshold correction is needed to realize the observed value $M_h=125.25\,\mathrm{GeV}$.  In the present case, we checked that, with a moderate choice of $m_{\tilde{t}} \lesssim 100\,\mathrm{TeV}$ (and thus $m_{\tilde{t}} \sim \mu$), the observed Higgs mass can be realized in the parameter region consistent with the vacuum stability constraint, which will be shown in subsequent sections. (We note, however, that a larger value of the stop mass may be required in the region excluded by the stability of the EW vacuum. However, it does not affect the upper bound on the smuon mass, which we will derive later from the stability of the EW vacuum.)

In between the two matching scales, we solve the RG equations of the corresponding EFT.
For running of the SM parameters, we use the two-loop RG equations \cite{Luo:2002ey} augmented by important three-loop contributions calculated in \cite{Buttazzo:2013uya}.
Also, the new physics contributions to the RG equations and the beta functions of new couplings in the slepton EFT are calculated at the one-loop level.
We summarize these contributions in Appendix \ref{sec:beta}.
Since all the SM parameters are fixed at $Q=M_t$ or below, while the other couplings are determined at $Q=M_S$, we solve the RG evolution in $M_t < Q < M_S$ iteratively to obtain the consistent solution.

\subsection{SUSY contribution to the muon $g-2$: calculation in EFT}

Now, we explain how we calculate $a_\mu^{\rm (SUSY)}$, the SUSY
contribution to the muon $g-2$. The EFT parameters introduced in previous subsections
are used to calculate $a_\mu^{\rm (SUSY)}$.

The mass matrix of the smuons is given by
\begin{align}
  {\bf M}^2_{\tilde{\mu}} =
  \left( \begin{array}{cc}
    m_{LL}^2 & - T_2 v
    \\
    - T_2 v & m_{RR}^2
  \end{array} \right),
\end{align}
where
\begin{align}
  m_{LL}^2 \equiv m_{L2}^2 + (\lambda_{HL2}+\kappa_2) v^2,~~~
  m_{RR}^2 \equiv m_{R2}^2 + \lambda_{HR2} v^2,
\end{align}
with $v\simeq 174\ {\rm GeV}$ being the vacuum expectation value of the
SM-like Higgs.  The mass matrix can be diagonalized by a $2\times 2$
unitary matrix $U$ as
\begin{align}
  \mbox{diag} (m^2_{\tilde{\mu}_1}, m^2_{\tilde{\mu}_2}) =
  U^\dagger {\bf M}^2_{\tilde{\mu}} U,
\end{align}
where $m_{\tilde{\mu}_1}$ and $m_{\tilde{\mu}_2}$ are lighter and
heavier smuon masses, respectively.  The gauge eigenstates are related
to the mass eigenstates, denoted as $\tilde{\mu}_A$ ($A=1$, $2$), as
\begin{align}
  \left( \begin{array}{c}
    \tilde{\mu}_L \\     \tilde{\mu}_R
  \end{array} \right) = 
  \left( \begin{array}{cc}
    U_{L,1} & U_{L,2}
    \\
    U_{R,1} & U_{R,2}
  \end{array} \right)
  \left( \begin{array}{c}
    \tilde{\mu}_1 \\ \tilde{\mu}_2
  \end{array} \right)
  \equiv
  U 
  \left( \begin{array}{c}
    \tilde{\mu}_1 \\ \tilde{\mu}_2
  \end{array} \right).
\end{align}

At the one-loop level, the Bino-smuon loop contribution to the muon
anomalous magnetic moment is given by \cite{Moroi:1995yh}
\begin{align}
  a_\mu^{(\tilde{B}\tilde{\mu},\, 1\mathhyphen{\rm loop})} = 
  \frac{m_\mu^2}{16\pi^2}
  \sum_{A=1}^2 
  \frac{1}{m_{\tilde{\mu}_A}^2}
  \left[
    - \frac{1}{12} \mathcal{A}_A f_1 (x_A) 
    - \frac{1}{3} \mathcal{B}_A f_2 (x_A)
    \right],
\end{align}
where $x_A\equiv M_1^2/m_{\tilde{\mu}_A}^2$,
\begin{align}
  \mathcal{A}_A \equiv Y_L^2 U_{L,A}^2 + Y_R^2 U_{R,A}^2,~~~
  \mathcal{B}_A \equiv \frac{M_1 Y_L Y_R U_{L,A} U_{R,A}}{m_\mu},
\end{align}
and the loop functions are given by
\begin{align}
  f_1 (x) \equiv &\, 
  \frac{2}{(1-x)^4} (1 - 6x + 3x^2 + 2x^3 - 6x^2 \ln x),
  \\
  f_2 (x) \equiv &\, \frac{3}{(1-x)^3} (1 - x^2 + 2 x \ln x).
\end{align}
We also take into account the leading-order Higgsino contribution because the Higgsino mass may become as light as $\sim O(1)$ TeV when we consider the case of relatively light staus.  We include the leading-order
Bino-Higgsino-smuon diagram contributions to $a_\mu^{\rm (SUSY)}$
\cite{Moroi:1995yh}:
\begin{align}
  a_\mu^{(\tilde{B}\tilde{H}\tilde{\mu},\, {\rm LO})} = &\,
  \frac{g_1^2 m_\mu^2 M_1 \mu \tan\beta}{16\pi^2}
  \left[ -\frac{1}{20 m_{LL}^4} g (x_{GL}, x_{HL}) 
    + \frac{1}{10 m_{RR}^4} g (x_{GR}, x_{HR}) 
    \right],
\end{align}
where 
\begin{align}
  x_{GL} = \frac{M_1^2}{m_{LL}^2}, ~~~
  x_{HL} = \frac{\mu^2}{m_{LL}^2}, ~~~
  x_{GR} = \frac{M_1^2}{m_{RR}^2}, ~~~
  x_{HR} = \frac{\mu^2}{m_{RR}^2},
\end{align}
and
\begin{align}
  g (x_G, x_H) \equiv
  \frac{6(x_G+x_H+x_G x_H -3)}{(x_G-1)^2 (x_H-1)^2}
  + \frac{12}{x_G-x_H}
  \left[ \frac{x_G\ln x_G}{(x_G-1)^3}
  - \frac{x_H\ln x_H}{(x_H-1)^3} \right].
\end{align}
Because we consider the case where  Winos are much heavier than the
EW scale, diagrams containing Winos are neglected.

In the MSSM, some of the two-loop contributions to the muon anomalous
magnetic moment may become sizable.  One important contribution is
from the non-holomorphic correction to the muon Yukawa coupling
constant \cite{Marchetti:2008hw, Girrbach:2009uy}.  In the limit of
large $\tan\beta$ (i.e., large $T_2$ parameter), such an effect can be
significant.  In the present setup, it is taken into account when the EFT parameters are matched to the MSSM
parameters at the SUSY scale (see the discussion in the previous
subsection).  Another is the photonic two-loop correction
\cite{Degrassi:1998es, vonWeitershausen:2010zr}.  It
includes large QED logarithms and can affect the SUSY contribution to
the muon $g-2$ by $\sim 10\ \%$ or more.  The full photonic two-loop
correction relevant to our analysis is given by
\cite{vonWeitershausen:2010zr}
\begin{align}
  a_\mu^{({\rm SUSY,\, photonic})} = & \,
  \frac{m_\mu^2}{16\pi^2} \frac{\alpha}{4\pi} 
  \sum_{A=1}^2 \frac{1}{m_{\tilde{\mu}_A}^2}
  \Bigg[
    16
    \left\{
    - \frac{1}{12} \mathcal{A}_A f_1 (x_A) 
    - \frac{1}{3} \mathcal{B}_A f_2 (x_A)
    \right\} \ln \frac{m_\mu}{m_{\tilde{\mu}_A}}
    \nonumber \\ & \,
    - \left\{
    - \frac{35}{75} \mathcal{A}_A f_3 (x_A) 
    - \frac{16}{9} \mathcal{B}_A f_4 (x_A)
    \right\}
    + \frac{1}{4} \mathcal{A}_A f_1 (x_A) 
    \ln \frac{m_{\tilde{\mu}_A}^2}{Q_{\rm DREG}^2}
    \Bigg],
\end{align}
where $\alpha$ is the fine structure constant, $Q_{\rm DREG}$ is the
dimensional-regularization scale, and
\begin{align}
  f_3 (x) \equiv &\, \frac{4}{105(1-x)^4} 
  [
    (1-x) (-97x^2 -529x +2)
    + 6 x^2 (13x + 81) \ln x
    \nonumber \\ &\,
    +108x (7x + 4) \mbox{Li}_2 (1-x)
  ],
  \\
  f_4 (x) \equiv &\, \frac{-9}{4(1-x)^3}
  [
    (1+3) (x \ln x +x -1)
    + (6x+2) \mbox{Li}_2 (1-x)
  ].
\end{align}

In our analysis, the SUSY contribution to the muon anomalous magnetic
moment is evaluated as
\begin{align}
  a_\mu^{\rm (SUSY)} = 
  a_\mu^{(\tilde{B}\tilde{\mu},\, 1\mathhyphen{\rm loop})}
  + a_\mu^{(\tilde{B}\tilde{H}\tilde{\mu},\, {\rm LO})} 
  + a_\mu^{({\rm SUSY,\, photonic})}.
\end{align}

\section{Decay of the EW Vacuum}
\label{sec:vacuumdecay}
\setcounter{equation}{0}

Now, we present the detailed formulas necessary to perform the
one-loop calculation of the decay rate of the EW vacuum in the MSSM,
adopting the method developed in Refs.\ \cite{Coleman:1977py, Callan:1977pt,
  Coleman:1985rnk}.  In order to study the implications of the muon
$g-2$ anomaly in the MSSM, we concentrate on the effects of sleptons.
In particular, because we are interested in the upper bound on the
smuon masses in the parameter region where the muon $g-2$ anomaly is
ameliorated by the SUSY contribution, we consider the case that the
smuon masses can be maximized for a given value of $a_\mu^{\rm
  (SUSY)}$.  In such a case, the trilinear scalar couplings of the
sleptons are enhanced, which may result in the instability of the EW
vacuum.

The decay rate of the false vacuum per unit volume (called ``the bubble
nucleation rate'') is expressed in the following form:
\begin{equation}
    \gamma=\mathcal A e^{-\mathcal B},
\end{equation}
where $\mathcal B$ is the bounce action while $\mathcal A$ is a
prefactor having mass-dimension four. The prefactor $\mathcal A$ is
obtained by integrating out the fluctuations around the bounce
configuration.  It has been often the case that $\mathcal A$ is simply
estimated as $\sim \Lambda^4$, where $\Lambda$ is a typical mass scale
in association with the bounce configuration. However, it has been
pointed out that, with explicitly integrating out the fluctuations,
$\ln\mathcal A/\Lambda^4$ and ${\mathcal B}$ can be of the same order
\cite{Endo:2015ixx}. Thus the calculation of $\mathcal A$ is important
for the accurate determination of the decay rate.

In gauge theories, special care is necessary to calculate 
${\cal A}$ because it should be performed with maintaining the gauge
invariance.  A prescription of the gauge invariant calculation of
${\cal A}$ is given in Refs.\  \cite{Endo:2017gal, Endo:2017tsz} for the
single-field bounce.  For the case of the decay rate of the EW vacuum
in the SM, the prefactor was first evaluated in \cite{Isidori:2001bm}
and then reevaluated with the correct treatment of the gauge degrees
of freedom in
\cite{Andreassen:2017rzq,Chigusa:2017dux,Chigusa:2018uuj} using the
results of \cite{Endo:2017gal, Endo:2017tsz}.  Then, the prescription
is generalized to a multi-field bounce in Ref.\ \cite{Chigusa:2020jbn},
which enabled the precise calculation of decay rates in more complex
setups.

In this section, we explain how the decay rate of the EW vacuum can be
studied in the MSSM using the prescription given in Ref.\
\cite{Chigusa:2020jbn}.  We calculate the decay rate using the EFT
defined in the previous section. Thus, in this section, all the
coupling constants are understood as those of the EFT.

\subsection{Bounce}

The first step to calculate the decay rate of a false vacuum is to
determine the bounce.  The bounce is an
$O(4)$-symmetric solution of the equations of motion derived from four-dimensional (4D) Euclidean field theory.  In the present case, there are several scalar fields contributing to the bounce, i.e., Higgs boson and sleptons.
In the following, we consider two cases:
\begin{itemize}
\item[(i)] The case that smuons are the only sfermions that may
  affect the stability of the EW vacuum; selectrons and staus are
  assumed to be so heavy that they are irrelevant.  In this case, we consider the EFT containing only the Bino and the smuons (as well as SM particles), and study the
  instability induced by $T_2$.  Then, we consider the bounce
  configuration parameterized as follows:
  \begin{align}
    H=\frac{1}{\sqrt{2}}\mqty(0\\\rho_h(r)),~~~
    \tilde \ell_{L2}=\frac{1}{\sqrt{2}}\mqty(0\\\rho_L(r)),~~~
    \tilde \ell_{R2}=\frac{1}{\sqrt{2}}\rho_R(r),
  \end{align}
  where $\rho_I$'s ($I=h$, $L$, and $R$) are functions that depend
  only on the Euclidean radius $r$.
\item[(ii)] The case that all the sleptons are relatively light so
  that we consider the EFT containing three generations of sleptons.  In this case, because $T_3$ is
  the largest among $T_\alpha$'s, staus play the most important role for the stability of the EW vacuum.  Thus, we concentrate
  on the bounce configuration consisting of staus as well as $H$,
  parameterized as follows:
  \begin{align}
    H=\frac{1}{\sqrt{2}}\mqty(0\\\rho_h(r)),~~~
    \tilde \ell_{L3}=\frac{1}{\sqrt{2}}\mqty(0\\\rho_L(r)),~~~
    \tilde \ell_{R3}=\frac{1}{\sqrt{2}}\rho_R(r).
  \end{align}
\end{itemize}
Notice that the configuration of the Higgs field is fixed by using the
local $SU(2)_L\times U(1)_Y$ transformation, while the directions of
$\tilde \ell_{L2}$ and $\tilde \ell_{L3}$ are chosen so that the
potential is destabilized by the trilinear scalar couplings.

The bounce configuration is obtained by solving the Euclidean equations of
motion:
\begin{equation}
  \partial_r^2\rho_I+\frac{3}{r}\partial_r\rho_I=\frac{dV}{d\rho_I},
\end{equation}
imposing the boundary conditions given by
\begin{align}
    \rho'_I(0)=0,~~~
    \rho_h(\infty)&=v_h,~~~
    \rho_L(\infty)=\rho_R(\infty)=0,
\end{align}
where the ``prime'' denotes the derivative with respect to $r$. Here,
$v_h$ is the Higgs amplitude at the local minimum of the scalar
potential in the EFT; because the Higgs mass parameter and the quartic
coupling in the SM are different from those in the EFT (see
Eqs.\ \eqref{eq:dellH} and \eqref{eq:delmHSq}), $v_h\neq v$ in general.  With the bounce configuration being given, the
exponential suppression factor is given by
\begin{equation}
    \mathcal B=S_E^{\rm (bounce)}-S_E^{\rm (false)},
\end{equation}
where $S_E^{\rm (false)}$ is the Euclidean action at the false vacuum and $S_E^{\rm (bounce)}$ is that of the bounce.

In our numerical calculation, we use the gradient flow method \cite{Chigusa:2019wxb,
  Sato:2019axv} to obtain the bounce configuration.  The details are given in Appendix \ref{sec:bounce}.

\subsection{Prefactor}

The prefactor $\mathcal{A}$ includes quantum corrections to $\mathcal B$ and it is important to evaluate both $\mathcal A$ and $\mathcal B$ to calculate the decay rate of the EW vacuum accurately.  The prefactor can be expressed as
\begin{equation}
  \mathcal A=2\pi\mathcal J_{\rm EM}\frac{\mathcal B^2}{4\pi^2}
  \mathcal A^{(A_\mu\varphi c\bar c)}
  \mathcal A^{(\psi)},
\end{equation}
where $\mathcal{A}^{(A_\mu\varphi c\bar c)}$ is the contribution of gauge bosons
and Faddeev-Popov ghosts as well as scalars, while
$\mathcal{A}^{(\psi)}$ is due to fermions.  Here, $2\pi$ is the
volume of the global $U(1)_{\rm EM}$ symmetry which is broken by the
bounce configuration, and $\mathcal J_{\rm EM}$ is the Jacobian in
association with the symmetry breaking (see Appendix
\ref{sec:fluc_op}).

At the one-loop level, the prefactor is obtained by evaluating the functional determinants of fluctuation operators. Here, the fluctuation operators are defined through $\mathcal M^{(X)}=\delta^2 S_E/\delta X_i\delta X_j$, where $S_E$ is the Euclidean action and $X_i$ denotes a field in the model. The functional determinants should be  evaluated around the bounce and the false vacuum to calculate the decay rate of the false vacuum. We utilize the $O(4)$ symmetry of the bounce to decompose the fluctuation operators into radial ones.
Each radial operator is labeled by ``the angular momentum'' $\ell$, which runs over integers. The contributions $\mathcal A^{(A_\mu\varphi c\bar c)}$ and $\mathcal A^{(\psi)}$ are expressed as
\begin{align}
  \mathcal A^{(A_\mu\varphi c\bar c)} = &\, \frac{\det\mathcal M_0^{(c\bar c)}}{\det\mathcal {\widehat M}_0^{(c\bar c)}}\qty(\frac{\det'\mathcal M_0^{(S\varphi)}}{\det\mathcal {\widehat M}_0^{(S\varphi)}})^{-1/2}\qty(\frac{\det'\mathcal M_1^{(SL\varphi)}}{\det\mathcal {\widehat M}_1^{(SL\varphi)}})^{-2}\prod_{\ell=2}^\infty\qty(\frac{\det\mathcal M_\ell^{(SL\varphi)}}{\det\mathcal {\widehat M}_\ell^{(SL\varphi)}})^{-\frac{(\ell+1)^2}{2}},
\\
  \mathcal A^{(\psi)} = &\,\prod_{\ell=0}^\infty\qty(\frac{\det\mathcal M_\ell^{(\psi)}}{\det\mathcal {\widehat M}_\ell^{(\psi)}})^{\frac{(\ell+1)(\ell+2)}{2}},
\end{align}
where $\mathcal M_\ell^{(X)}$'s indicate radial fluctuation operators around the bounce and $\widehat{\mathcal M}_\ell^{(X)}$'s are those around the false vacuum. Here, $S$ and $L$ in the superscript denote the specific modes in the gauge fluctuations, and each operator is explicitly given in Appendix \ref{sec:fluc_op}.
For $\ell=0,1$, there appear a gauge zero mode and translational zero modes in association with the spontaneous breaking of the $U(1)_{\rm EM}$ symmetry and the translations; $\det'$ denotes the functional determinant after the zero mode subtraction.

The radial fluctuation operators have the form of
\begin{align}
  \mathcal M_\ell =
  -\partial_r^2-\frac{3}{r}\partial_r+\frac{L^2_\ell}{r^2}+m^2(r),
\end{align}
and 
\begin{align}
  \widehat{\mathcal M}_\ell = 
  -\partial_r^2-\frac{3}{r}\partial_r+\frac{L^2_\ell}{r^2}+\widehat m^2,
\end{align}
where $L^2_\ell$, $m^2(r)$ and $\widehat m^2$ are $n\times n$ matrices with $n$ being an integer that depends on the operator. Here, $L^2_\ell$ is a diagonal matrix with elements being the eigenvalues of $r^2\partial^2$ acting on functions depending only on angular variables, and it depends only on $\ell$. These fluctuation operators can be block-diagonalized and each block is given in Appendix \ref{sec:fluc_op}.

Using the method given in Refs.\ \cite{Gelfand:1959nq, Dashen:1974ci, Coleman:1985rnk, Kirsten:2003py, Endo:2017tsz}, the functional
determinants can be evaluated as
\begin{equation}
    \frac{\det\mathcal M_\ell}{\det\widehat{\mathcal M}_\ell}=\lim_{r\to\infty}\frac{\det\psi_\ell(r)}{\det\widehat\psi_\ell(r)},\label{eq_the_theorem}
\end{equation}
where $\psi_\ell(r)$ and $\widehat\psi_\ell(r)$ are $n\times n$ functions satisfying
\begin{align}
  \qty[-\partial_r^2-\frac{3}{r}\partial_r+\frac{L^2_\ell}{r^2}+m^2]\psi_\ell(r)&=0,
  \label{diffeq_psi}
\end{align}
and 
\begin{align}
  \qty[-\partial_r^2-\frac{3}{r}\partial_r+\frac{L^2_\ell}{r^2}+\widehat m^2]\widehat\psi_\ell(r)&=0,
  \label{diffeq_hatpsi}
\end{align}
with
\begin{equation}
    \lim_{r\to0}\frac{\det\psi_\ell(r)}{\det\widehat\psi_\ell(r)}=1.
\end{equation}

Since the $U(1)_{\rm EM}$ and translation symmetries are broken by the bounce, there appear zero modes.
The zero modes can be subtracted from the functional determinants as
\begin{equation}
    \frac{\det'\mathcal M_\ell}{\det\widehat{\mathcal M}_\ell}=\lim_{\nu\to0}\frac{1}{\nu}\frac{\det[\mathcal M_\ell+\nu]}{\det\widehat{\mathcal M}_\ell}.
\end{equation}
From Eq.~\eqref{eq_the_theorem}, one can see that this corresponds to
\begin{equation}
    \frac{\det'\mathcal M_\ell}{\det\widehat{\mathcal M}_\ell}=\lim_{\nu\to0}\frac{1}{\nu}\lim_{r\to\infty}\frac{\det(\psi_\ell(r)+\nu\check\psi_\ell(r))}{\det\widehat\psi_\ell(r)},
\end{equation}
where $\check\psi$ is the function satisfying 
\begin{align}
    \qty[-\partial_r^2-\frac{3}{r}\partial_r+\frac{L^2_\ell}{r^2}+m^2]\check\psi_\ell(r)&=-\psi_\ell(r),
\end{align}
with $\check\psi_\ell(0)=0$.

Although the contribution from each partial wave is finite, the prefactor diverges after taking into account all the contributions.
The ratio of the functional determinants can be interpreted as one-loop bubble diagrams with insertions of $(m^2-\widehat m^2)$; the divergences exist only in the diagrams with one and two insertions. This motivates us to consider a regularized quantity:
\begin{align}
    \ln\frac{\det\mathcal M_\ell}{\det\widehat{\mathcal M}_\ell}-s_\ell&=\lim_{r\to\infty}\qty[\ln\det\widehat\psi_\ell^{-1}\psi_\ell-\tr\widehat\psi_\ell^{-1}\psi_\ell^{(1)}-\tr\widehat\psi_\ell^{-1}\psi_\ell^{(2)}+\frac12\tr\widehat\psi_\ell^{-1}\psi_\ell^{(1)}\widehat\psi_\ell^{-1}\psi_\ell^{(1)}],
\end{align}
where the functions $\psi^{(1)}$ and $\psi^{(2)}$ are solutions of the following equations:
\begin{align}
    \qty[-\partial_r^2-\frac{3}{r}\partial_r+\frac{L^2_\ell}{r^2}+\widehat m^2]\psi_\ell^{(1)}&=-(m^2-\widehat m^2)\widehat\psi_\ell,\\
    \qty[-\partial_r^2-\frac{3}{r}\partial_r+\frac{L^2_\ell}{r^2}+m^2]\psi_\ell^{(2)}&=-(m^2-\widehat m^2)\psi_\ell^{(1)}.
\end{align}
Here, $s_\ell$ denotes the divergent part of each partial wave.

The renormalized ratio of the functional determinants is obtained as
\begin{equation}
  \eval{\ln\frac{\det\mathcal M}{\det\widehat{\mathcal M}}}_{\rm \overline{MS}}=\sum_{\ell}d_\ell\qty(\ln\frac{\det\mathcal M_\ell}{\det\widehat{\mathcal M}_\ell}-s_\ell)+ s_{\rm \overline{MS}},\label{eq_prefactor_renormalization}
\end{equation}
where $d_\ell$ is the degeneracy of the partial waves, and 
\begin{equation}
    s_{\overline{\rm MS}}=\eval{\sum_\ell d_\ell s_\ell}_{\overline {\rm MS}},
\end{equation}
which is the finite quantity after the subtraction of the divergence via the $\overline{\rm MS}$ scheme.
Each term in the summation
in the right-hand side of Eq.\ \eqref{eq_prefactor_renormalization}
scales as $\sim\ell^{-2}$ at a large $\ell$ and hence the sum is convergent. We calculate them up to a large enough $\ell$ and extrapolate the results to $\ell=\infty$.
Since $s_\ell$ is defined through the expansion with respect to $(m^2-\widehat m^2)$, it can be calculated diagrammatically.
By the direct evaluation of loop integrals with the $\overline{\rm MS}$ scheme, we can obtain $s_{\rm \overline{MS}}$ (for more details, see Appendix \ref{sec:ct}).

\section{Numerical Results}
\label{sec:results}
\setcounter{equation}{0}

Now, we discuss the stability of the EW vacuum
in the MSSM in connection with the muon anomalous magnetic moment.  We
calculate the decay rate of the EW vacuum with taking into account the
loop effects due to the EW gauge bosons, SM-like Higgs boson, and top
quark as well as the Bino and the sleptons in the EFT.
(For our
numerical calculations, we use the
coupling constants at the renormalization scale of $Q=M_t$.)

In the following argument, we parameterize the bubble nucleation rate
as
\begin{align}
  \Seff \equiv - \ln \left( \frac{\gamma}{1\ {\rm GeV}^4} \right).
\end{align}
Then, in order for the bubble nucleation rate within the Hubble
volume, $\frac{4}{3}\pi H_0^{-3}$ (with $H_0$ being the Hubble
parameter of the present universe), to be smaller than $t_{\rm
  now}^{-1}$, $\Seff$ is constrained as
\begin{align}
  \Seff > 386.
  \label{seffbound}
\end{align}
Hereafter, we take the above constraint as a requirement for the
stability of the EW vacuum.

\subsection{Case with only second-generation sleptons and Bino}

We first consider the minimal case in which the Bino and the
second-generation sleptons are the only superparticles whose masses
are lighter than a few TeV.  Other superparticles are assumed to be so
heavy that they do not affect the lifetime of the EW vacuum.
Here, we extend the previous analysis \cite{Chigusa:2022xpq} and
investigate the dependence of the stability of the EW
vacuum on $a_\mu^{\rm (SUSY)}$.

\begin{figure}
  \centering
  \includegraphics[width=0.65\linewidth]{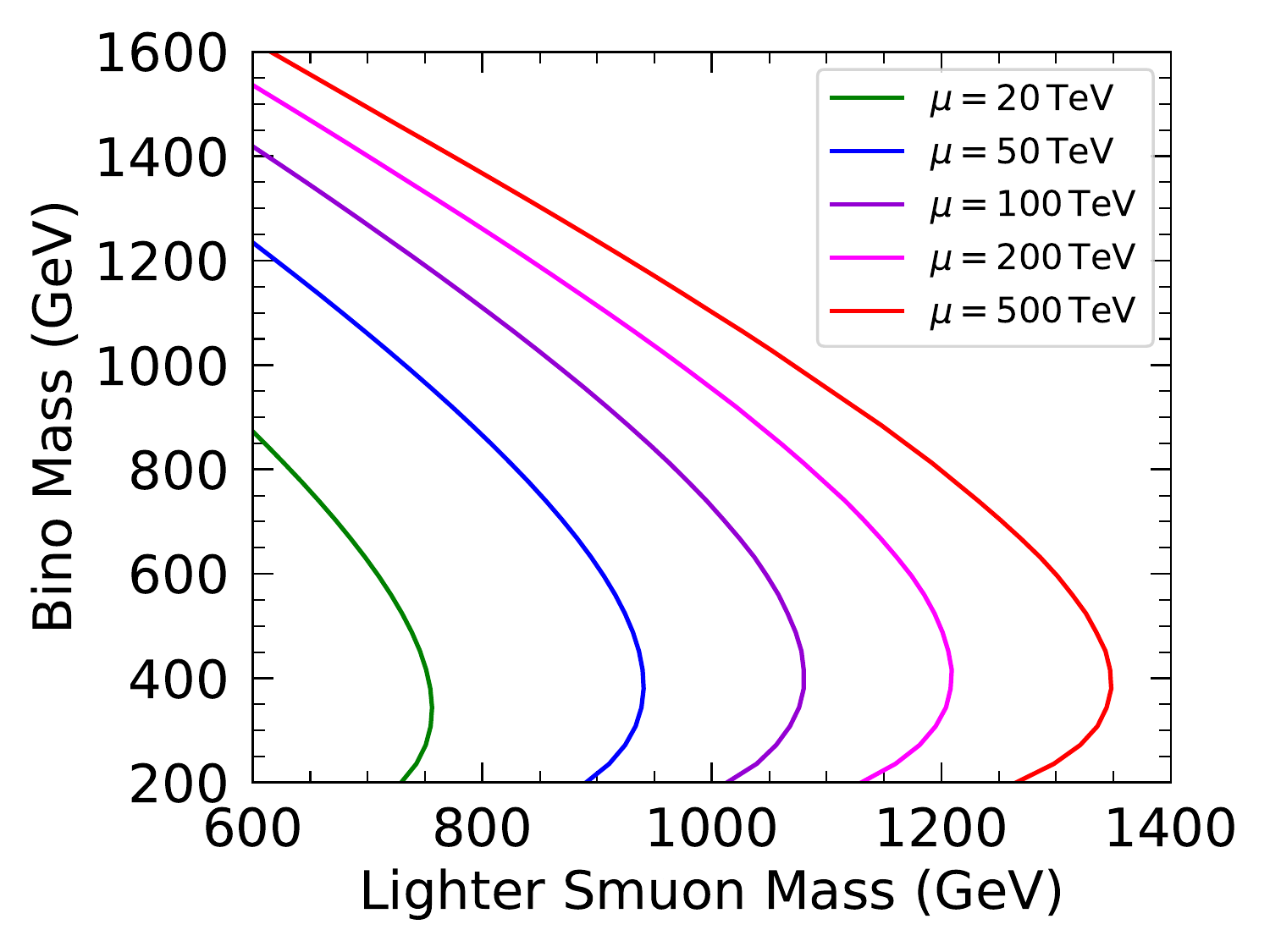}
  \caption{Contours of constant $\mu$ to realize $a_\mu^{\rm
      (SUSY)}=\Delta a_\mu^{(0\sigma)}$, taking $\tan\beta=50$, and
    $m_{L2}=m_{R2}$.  The green, blue, violet, magenta, and red
    contours are for $\mu=20$, $50$, $100$, $200$, and $500\ {\rm
      TeV}$, respectively.}
  \label{fig:muh_tb50_r10_s0}
\vspace{7mm}
  \centering
  \includegraphics[width=0.65\linewidth]{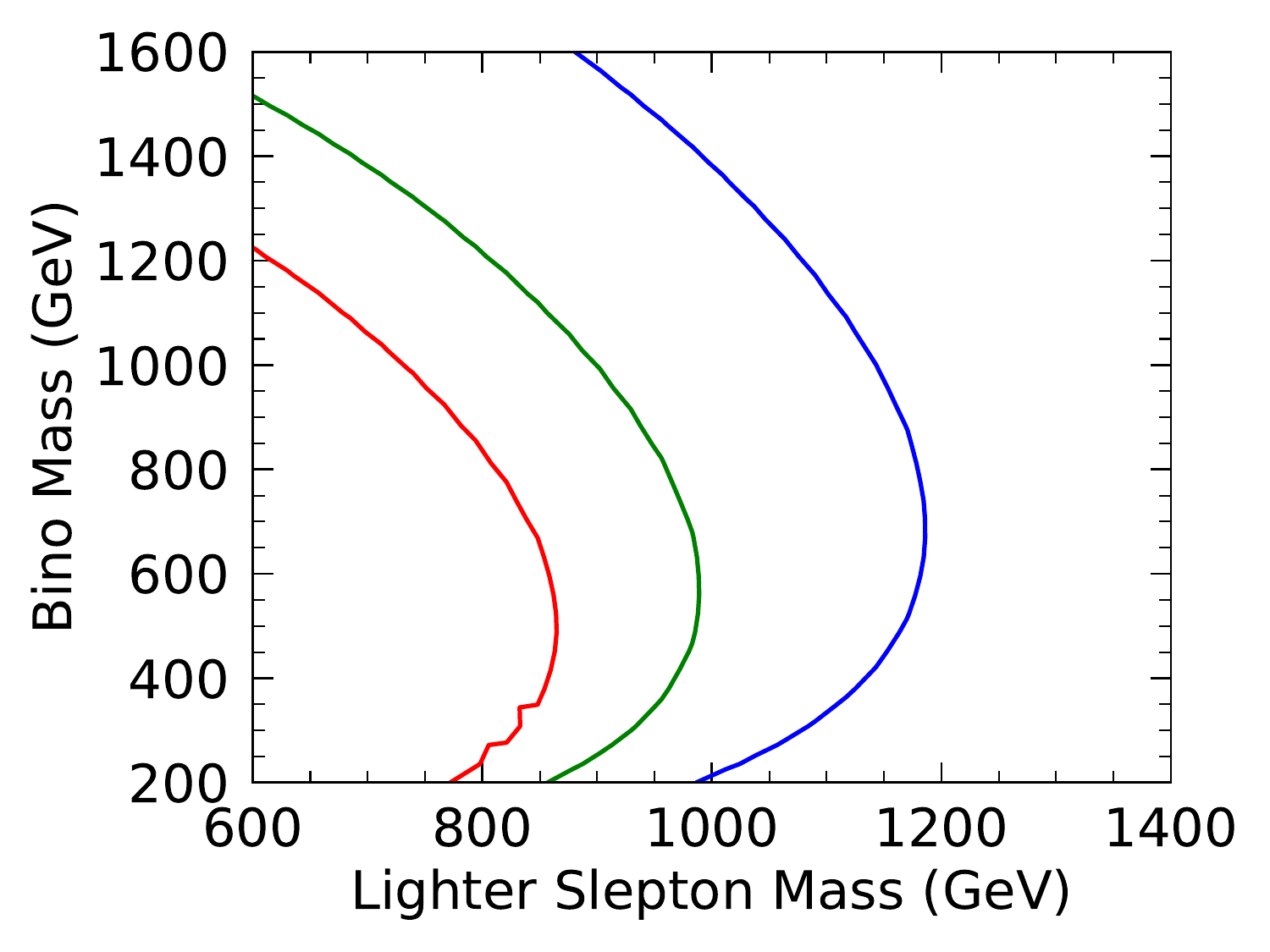}
  \caption{The boundary between the regions with and without the
    charge-breaking absolute minimum of the potential, taking
    $a_\mu^{\rm (SUSY)}=\Delta a_\mu^{(0\sigma)}$ (red), $\Delta
    a_\mu^{(1\sigma)}$ (green), and $\Delta a_\mu^{(2\sigma)}$ (blue).
    Here, we take $\tan\beta=50$, and $m_{L2}=m_{R2}$.}
  \label{fig:vmin_tb50_r10}
\end{figure}

For the case of only the smuons (as well as the muon sneutrino), the
trilinear coupling constant $T_2$ is important for the study of the
stability of the EW vacuum.  With $a_\mu^{\rm (SUSY)}$ being fixed,
$T_2$ is determined once other EFT parameters (like the Bino and smuon
masses) are given; $T_2$ becomes larger as the smuons become heavier.
This is due to the fact that the loop functions are suppressed as the
smuon mass becomes larger so that the left-right mixing should be
enhanced (see Fig.\ \ref{fig:feyndiag}).  For a given value of $T_2$,
the SUSY invariant Higgs mass parameter $\mu$ is fixed via
Eq.\ \eqref{eq:TBC}.  In Fig.\ \ref{fig:muh_tb50_r10_s0}, we show the
contours of constant $\mu$ on the lighter smuon mass vs.\ Bino mass
plane, assuming $a_\mu^{\rm (SUSY)}=\Delta a_\mu^{(0\sigma)}$; here,
we take $\tan\beta=50$ and $m_{L2}=m_{R2}$.  (Notice that $\mu$
becomes larger if we assume a smaller value of $\tan\beta$ than $50$;
with fixing $a_\mu^{\rm (SUSY)}$ as well as the smuon and Bino masses,
$\mu$ is approximately inversely proportional to $\tan\beta$.)  We can see that
$\mu$ is much larger than the EW scale, which justifies our analysis 
using the slepton EFT.

With the $\mu$ parameter being enhanced, there may appear a charge-breaking minimum of the potential at which sfermions (as well as
Higgses) acquire non-vanishing expectation values.  The energy density
of the charge breaking minimum is often smaller than that of the
EW vacuum; if so, the EW vacuum is not absolutely stable. In Fig.\ \ref{fig:vmin_tb50_r10}, on the
lighter smuon mass vs.\ Bino mass plane, we show the parameter region
in which the charge breaking minimum becomes the true vacuum; the
contours in the figure show the boundary between the regions with and
without the charge breaking absolute minimum.  Here we take
$\tan\beta=50$ and $m_{L2}=m_{R2}$; the red, green, and blue contours
are for $a_\mu^{\rm (SUSY)}=\Delta a_\mu^{(0\sigma)}$, $\Delta
a_\mu^{(1\sigma)}$, and $\Delta a_\mu^{(2\sigma)}$, respectively.  The
EW vacuum becomes unstable at the right-hand side of the
contours.  

\begin{figure}[t]
  \centering
  \includegraphics[width=0.65\linewidth]{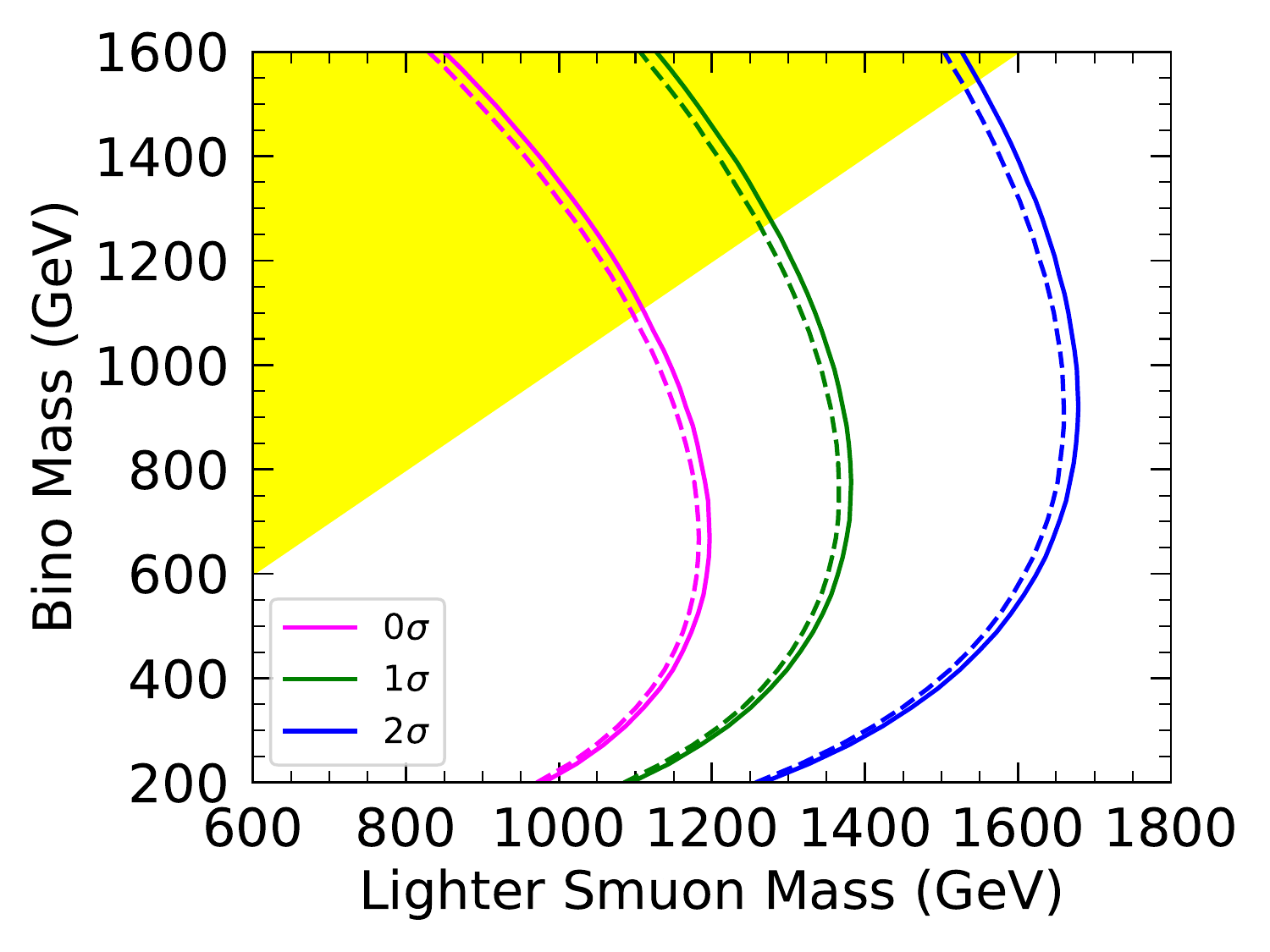}
  \caption{Contours of $S_{\rm eff}=387$, corresponding to
    $\gamma_{\rm EW}=t_{\rm now}^{-1}$, for the case of $m_{L2}=m_{R2}$.
    Magenta, green, and blue contours are for $a_\mu^{\rm
      (SUSY)}=\Delta a_\mu^{(0\sigma)}$, $\Delta a_\mu^{(1\sigma)}$,
    and $\Delta a_\mu^{(2\sigma)}$, respectively, with $\tan\beta=10$
    (solid) and $50$ (dashed). In the yellow-shaded region, the
    lighter smuon becomes lighter than the Bino.}
  \label{fig:seff387_smu}
\end{figure}

Even if the EW vacuum is unstable, we may still live on it if its
lifetime is (much) longer than the present cosmic age.  We calculate
the decay rate of the EW vacuum using the formulas given in the
previous section.  Then, we derive constraints on the parameter space.

In Fig.\ \ref{fig:seff387_smu}, we show the contours of $S_{\rm
  eff}=387$ on the lighter smuon mass vs.\ Bino mass plane, taking
$a_\mu^{\rm (SUSY)}=\Delta a_\mu^{(0\sigma)}$, $\Delta
a_\mu^{(1\sigma)}$, and $\Delta a_\mu^{(2\sigma)}$.  Here, we take
$m_{L2}=m_{R2}$ and $\tan\beta=10$ and $50$. The value of $S_{\rm
  eff}$ becomes smaller as the smuons become heavier; thus the
contours in the figures show the maximal possible value of the lighter
smuon mass for a given Bino mass.  Notice that, as discussed in
Ref.\ \cite{Chigusa:2022xpq}, the upper bound on the lighter smuon
mass becomes smaller as the ratio $m_{L2}/m_{R2}$ deviates from $1$;
thus the bound is obtained from the study of the case of
$m_{L2}=m_{R2}$.  In the figure, we show the region in which the
lighter smuon becomes lighter than the Bino.  In such a parameter
region, the lighter smuon becomes the lightest among the MSSM
particles and it may be the lightest superparticle (LSP).  The LSP
is stable assuming $R$-parity conservation.  If the lighter smuon
is the LSP and is stable, collider and cosmological constraints may
apply.  The LHC experiment excludes stable sleptons lighter than $\sim
430\ {\rm GeV}$ \cite{CMS:2016kce, ATLAS:2019gqq}.  In addition,
cosmologically, the existence of a new stable charged particle is disfavored
because it can be produced just after the hot big bang and survives
until today.  These constraints are, however, evaded if there exists a
superparticle lighter than the smuon; the examples include gravitino
(i.e., the superpartner of the graviton) and axino (i.e., superpartner
of the axion).  An $R$-parity violation is another possibility to
avoid the constraints.  Because these constraints are model dependent,
we do not take them into account in deriving the vacuum stability
bound on the smuon mass.

\begin{figure}
  \centering
  \includegraphics[width=0.65\linewidth]{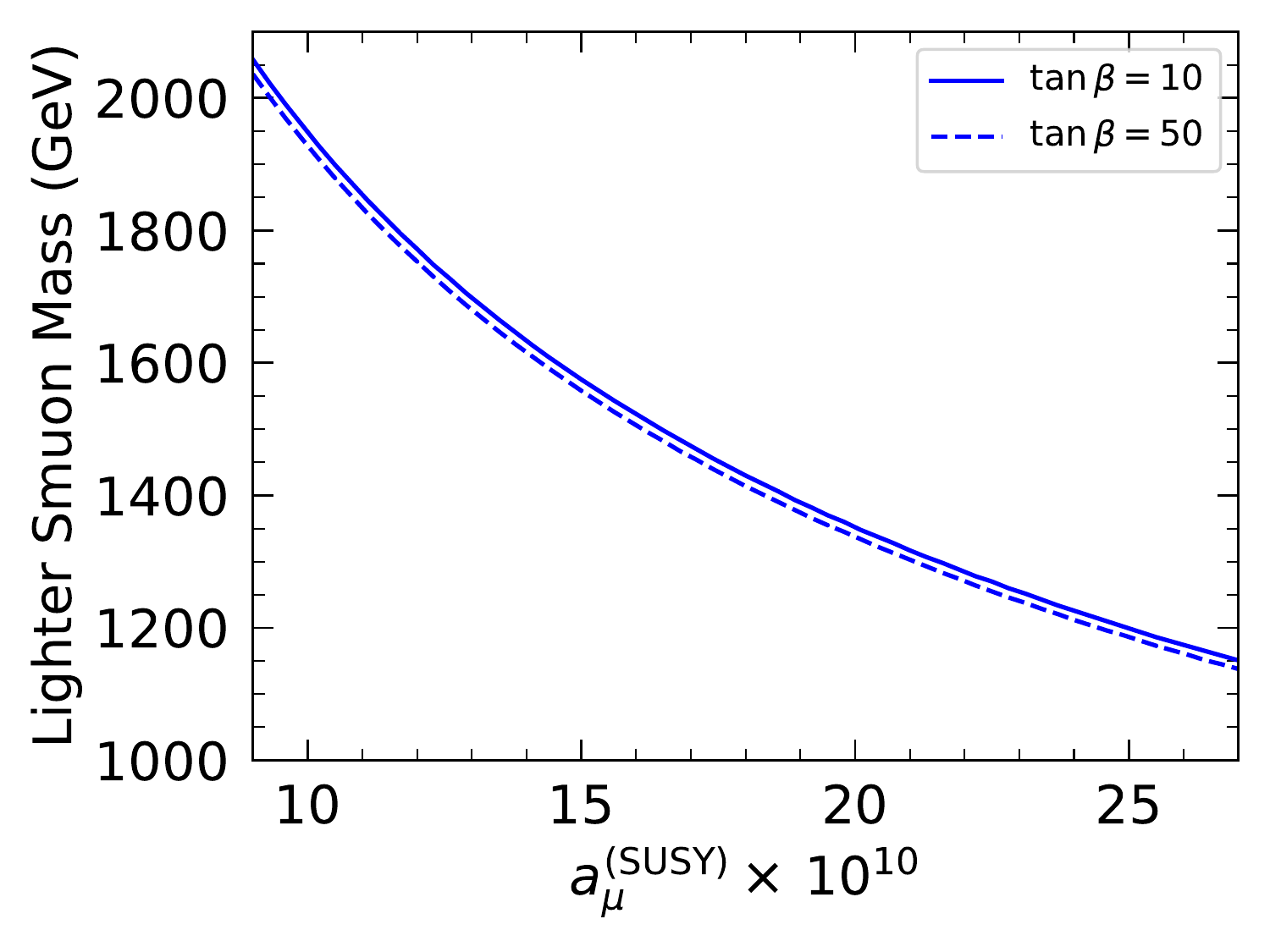}
  \caption{Maximal possible value of the lighter smuon mass as a
    function of $a_\mu^{\rm (SUSY)}$, taking $m_{L2}=m_{R2}$.  Solid
    and dashed lines are for $\tan\beta=10$ and $50$, respectively.}
  \label{fig:mslmax_da}
\end{figure}

The vacuum stability bound becomes severer as the value of
$a_\mu^{\rm (SUSY)}$ becomes larger.   In Fig.\ \ref{fig:mslmax_da}, we show the maximal
possible value of the lighter smuon mass as a function of $a_\mu^{\rm (SUSY)}$, taking $m_{L2}=m_{R2}$.  For the case of adopting the SM prediction
given in Eq.\ \eqref{amu(SM)}, with which the SM prediction deviates
from the experimental value by $4.2\sigma$, the constraint is
stringent; requiring $0\sigma$, $1\sigma$, and $2\sigma$ consistency,
the lighter smuon mass is required to be smaller than 
$1.18$, $1.37$, and $1.66\ {\rm TeV}$, respectively, for $\tan\beta=50$.  On the
contrary, if we adopt the lattice evaluation of the HVP contribution,
the bound on the smuon mass becomes less stringent.  For example, even
if we require that the SM prediction be equal to the central value of
$a_\mu^{\rm (exp)}$, the lighter smuon mass can be as heavy as
$1.86\ {\rm TeV}$ taking $a_\mu^{\rm (SM, BMW)}$ as SM prediction.

\subsection{Case with three generations of sleptons}

Next, we consider the case that all the slepton masses are comparable.
In such a case, the stau sector has the most serious effect on the
stability of the EW vacuum.

The value of $\mu$ necessary to solve the muon $g-2$ anomaly is
(almost) unchanged even with relatively light staus.  With the value
of $\mu$ suggested from the muon $g-2$ anomaly, the trilinear coupling
of staus may be significantly enhanced; it is an order of magnitude
larger than that of smuons because the trilinear couplings of
sfermions are proportional to corresponding Yukawa coupling constants
(as far as we can neglect the so-called $A$-terms, i.e., soft SUSY
breaking trilinear couplings).  Consequently, if the smuon and stau
masses are comparable, the EW vacuum is more easily
destabilized by the trilinear coupling of the staus.

We study the decay of the EW vacuum mediated by the bounce
configuration consisting of staus and Higgses.  For simplicity, we
assume that superparticles other than the sleptons and the Bino are much
heavier than the EW scale so that the decay rate of the EW vacuum can
be studied by the EFT containing only the sleptons and the Bino (as well as
the SM particles).

\begin{figure}
  \centering
  \includegraphics[width=0.65\linewidth]{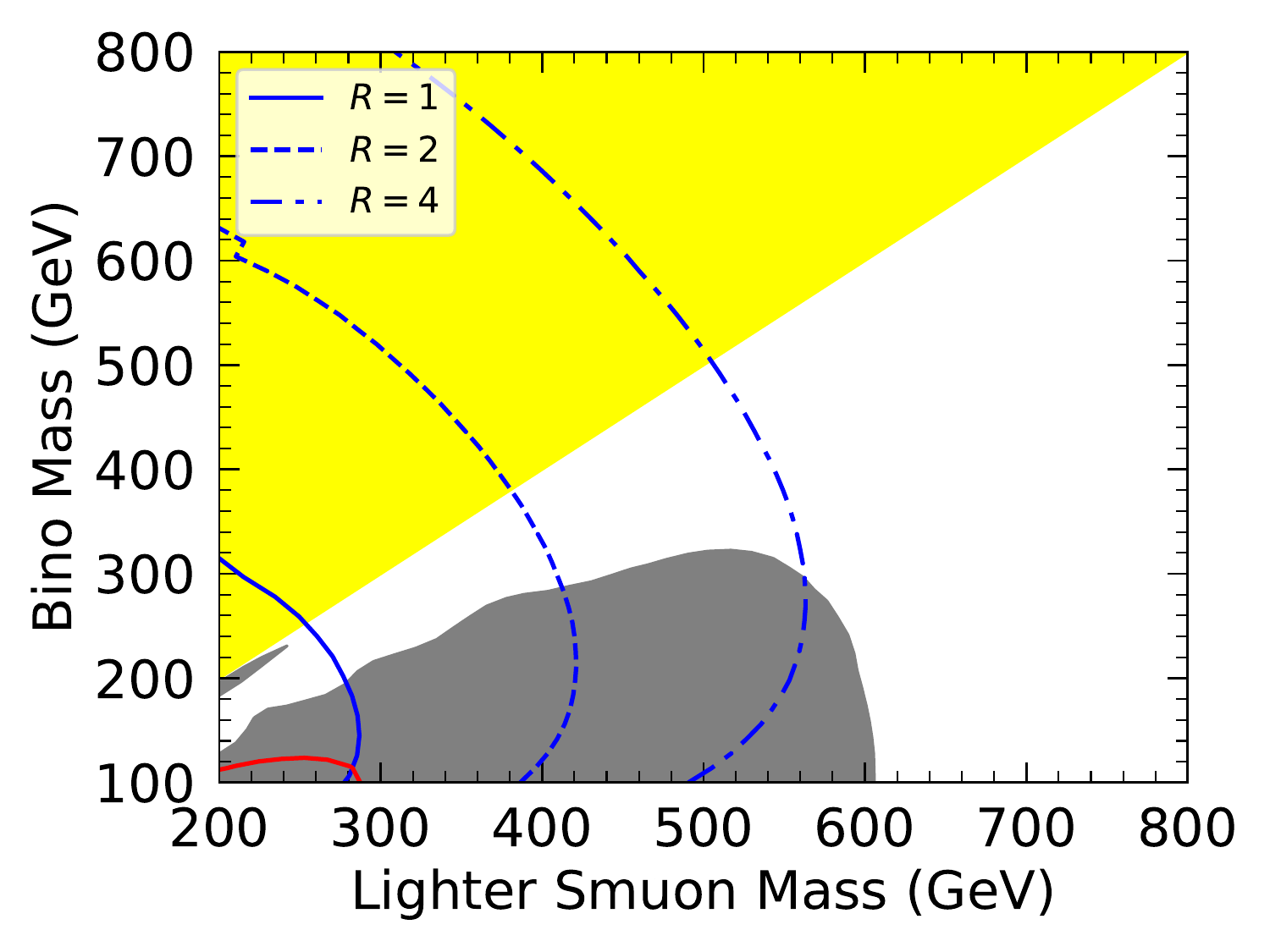}
  \caption{Contours of constant $S_{\rm eff}=387$ for $a_\mu^{\rm
      (SUSY)}=\Delta a_\mu^{(0\sigma)}$ with $\Rsmustau=1$ (solid),
    $2$ (dashed), and $4$ (dash-dotted), taking $\tan\beta=50$.  Here, we
    assume
    $m_{\tilde{e}_{L}}=m_{\tilde{e}_{R}}=m_{\tilde{\mu}_{L}}=m_{\tilde{\mu}_{R}}$
    and $m_{\tilde{\tau}_{L}}=m_{\tilde{\tau}_{R}}$, and parameterize
    the stau masses as $m_{\tilde{\tau}_{L,R}}=\Rsmustau\,
    m_{\tilde{\mu}_{L,R}}$.  The gray-shaded region is excluded by the
    ATLAS searches for the smuons assuming that the lightest
    neutralino is the LSP, while the lighter smuon becomes lighter
    than the Bino in the yellow-shaded region.  In addition, the red
    contour shows the boundary beyond which the lighter stau becomes
    the lightest MSSM particle. }
  \label{fig:seff387_stau_tb50_r10_s0}
\vspace{7mm}
  \includegraphics[width=0.65\linewidth]{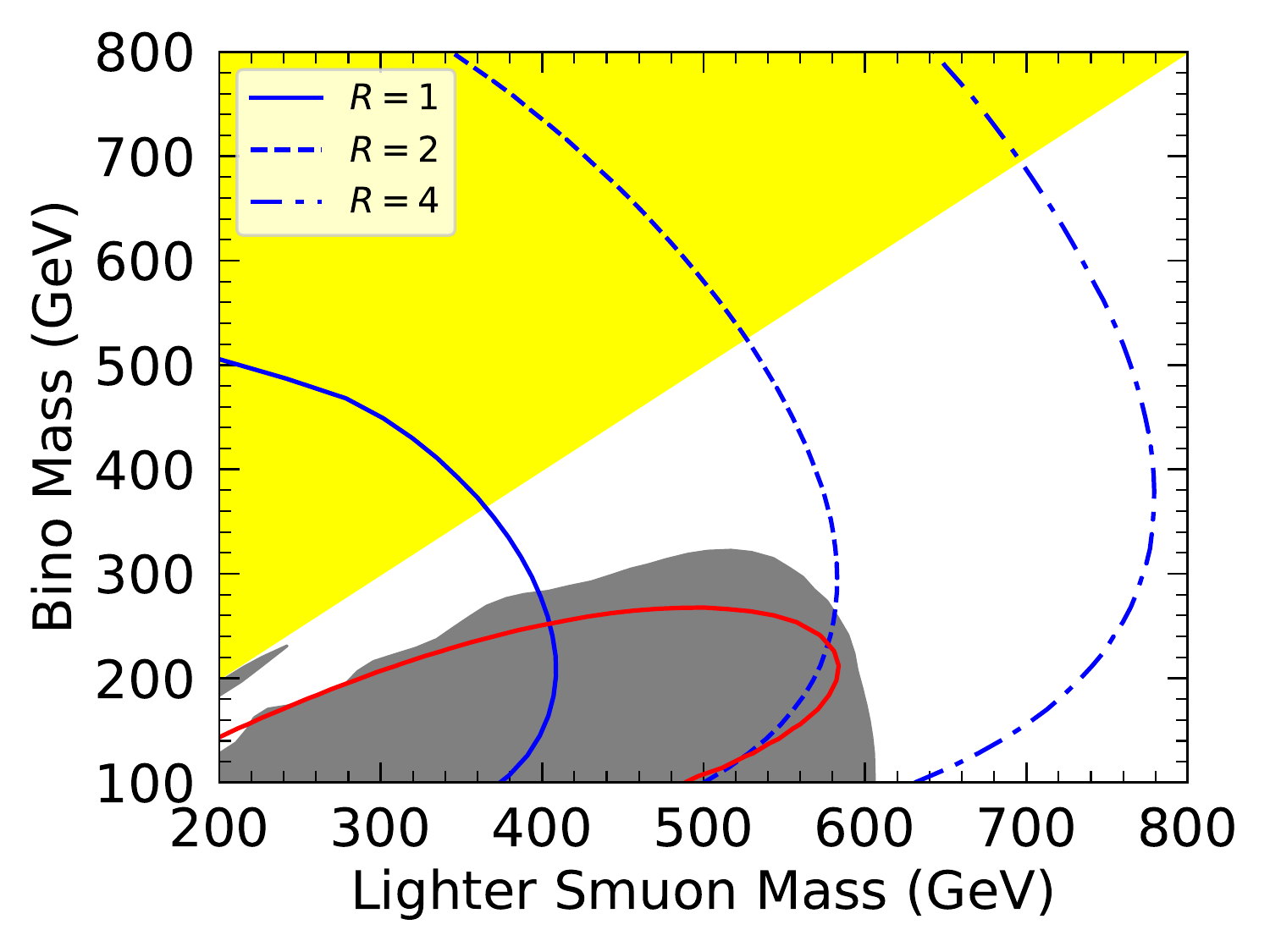}
  \caption{Same as Fig.\ \ref{fig:seff387_stau_tb50_r10_s0}, except
    for the case of $a_\mu^{\rm (SUSY)}=\Delta a_\mu^{(2\sigma)}$.}
  \label{fig:seff387_stau_tb50_r10_s2}
\end{figure}

In order to see how the upper bound on the lighter smuon mass depends on the mass hierarchy between the smuons and staus, we take $m_{\tilde{e}_{L}}=m_{\tilde{e}_{R}}=m_{\tilde{\mu}_{L}}=m_{\tilde{\mu}_{R}}$
and $m_{\tilde{\tau}_{L}}=m_{\tilde{\tau}_{R}}$, and parameterize the
stau masses as follows:
\begin{align}
  m_{\tilde{\tau}_{L,R}} = \Rsmustau\, m_{\tilde{\mu}_{L,R}},
\end{align}
where $\Rsmustau$ is a positive constant.  In
Figs.\ \ref{fig:seff387_stau_tb50_r10_s0} and
\ref{fig:seff387_stau_tb50_r10_s2}, we show the contours of constant
$S_{\rm eff}=387$ on the lighter smuon mass vs.\ Bino mass plane,
taking $\Rsmustau=1$, $2$ and $4$.  We consider the cases of
$a_\mu^{\rm (SUSY)}=\Delta a_\mu^{(0\sigma)}$ and $\Delta
a_\mu^{(2\sigma)}$ in Figs.\ \ref{fig:seff387_stau_tb50_r10_s0} and
\ref{fig:seff387_stau_tb50_r10_s2}, respectively.  

If all the slepton masses are of the same order, the lighter stau may become the LSP.  This is because the lighter stau
mass may be significantly reduced due to the large left-right stau
mixing.  In the present case, the lighter stau becomes the lightest
MSSM particle in a wide parameter space when $\Rsmustau$ becomes close
to $1$ (or smaller).  In Figs.\ \ref{fig:seff387_stau_tb50_r10_s0} and
\ref{fig:seff387_stau_tb50_r10_s2}, we show the contour beyond which
the lighter stau becomes the lightest MSSM particle for the case of
$\Rsmustau=1$ (red contour).  (For the cases of $\Rsmustau=2$ and $4$,
we have checked that the stau does not become the LSP in the parameter
region of our interest.)  In addition, as in the case of
Fig.\ \ref{fig:seff387_smu}, we show the region in which the lighter
smuon becomes lighter than the Bino (yellow-shaded region). We also
show the region which is excluded by the smuon searches by the ATLAS
experiment (gray region) \cite{ATLAS:2022rcw}. For $\Rsmustau=1$, the stau can be the LSP in a wide parameter region; such a parameter
region may conflict with the LHC constraint on the stable slepton
because the stable slepton lighter than $\sim 430\ {\rm GeV}$ is
excluded by the LHC \cite{CMS:2016kce, ATLAS:2019gqq}.  In addition,
cosmological constraints on stable charged particles may also apply.
However, as we have mentioned, these constraints are model
dependent and hence we do not take them into account to derive a
conservative bound.

\begin{figure}
  \centering
  \includegraphics[width=0.65\linewidth]{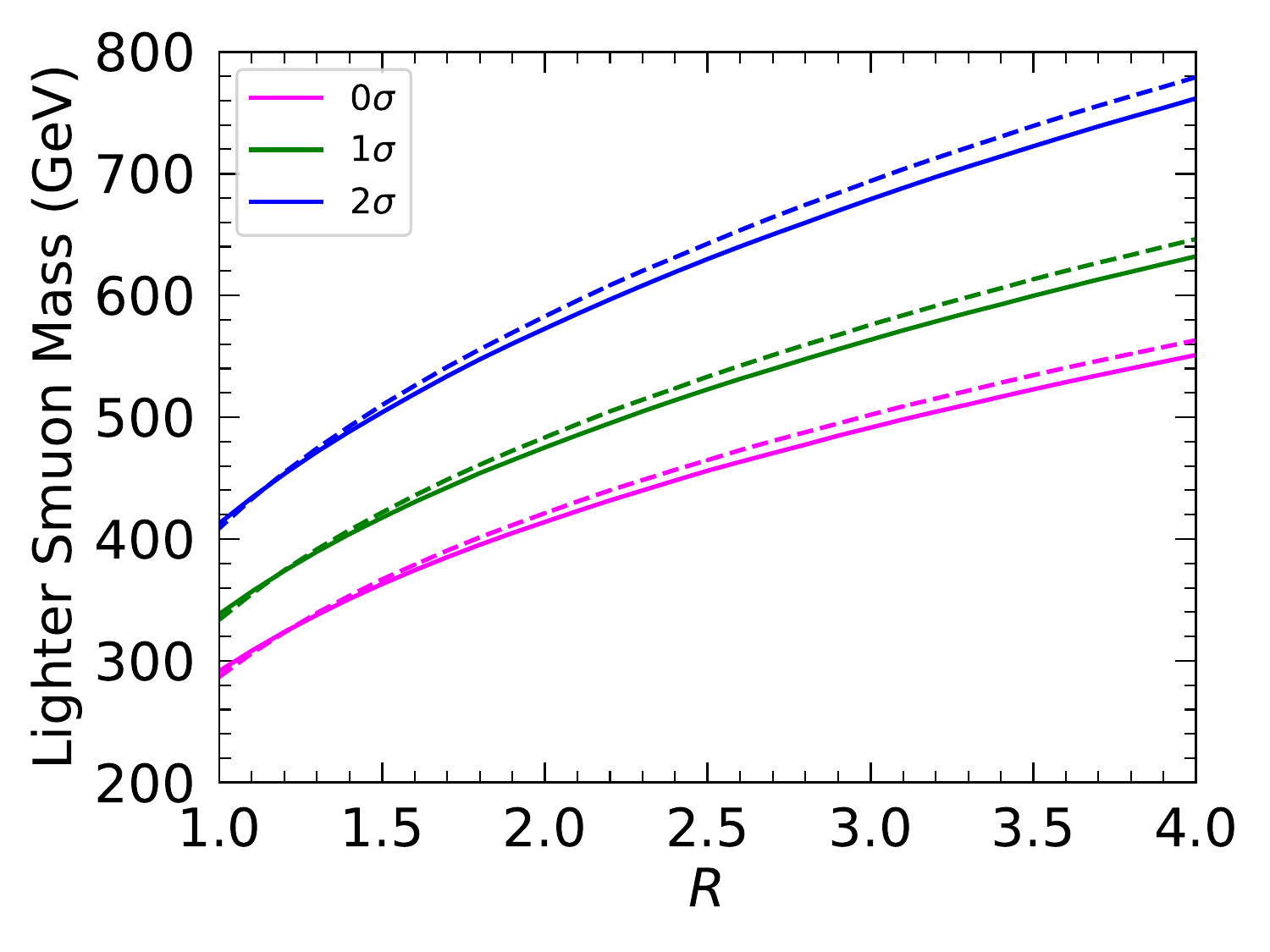}
  \caption{Upper bound on the lighter smuon mass as a function of
    $\Rsmustau$.  The magenta, green, and blue lines show the bound for
    $a_\mu^{\rm (SUSY)}=\Delta a_\mu^{(0\sigma)}$, $\Delta
    a_\mu^{(1\sigma)}$, and $\Delta a_\mu^{(2\sigma)}$, respectively,
    taking $\tan\beta=10$ (solid) and $50$ (dashed). Here, we
    assume
    $m_{\tilde{e}_{L}}=m_{\tilde{e}_{R}}=m_{\tilde{\mu}_{L}}=m_{\tilde{\mu}_{R}}$
    and $m_{\tilde{\tau}_{L}}=m_{\tilde{\tau}_{R}}$, and parameterize
    the stau masses as $m_{\tilde{\tau}_{L,R}}=\Rsmustau\,
    m_{\tilde{\mu}_{L,R}}$.}
  \label{fig:mslmax_stau_mrdep}
\end{figure}

To see how the upper bound depends on the ratio of the stau and smuon
masses, in Fig.\ \ref{fig:mslmax_stau_mrdep}, we show the
maximal possible value of the lighter smuon mass to guarantee the
longevity of the EW vacuum as a function of $\Rsmustau$; here we take
$a_\mu^{\rm (SUSY)}=\Delta a_\mu^{(0\sigma)}$ (magenta), $\Delta
a_\mu^{(1\sigma)}$ (green), and $\Delta a_\mu^{(2\sigma)}$ (blue). We
can see that the upper bound on the lighter smuon mass is
significantly lowered compared with the case only with the smuons.

\section{Conclusions and Discussion}
\label{sec:conclusions}
\setcounter{equation}{0}

In this paper, we have studied the stability of the EW vacuum in the
MSSM, paying particular attention to its relation to the SUSY
contribution to the muon anomalous magnetic moment, $a_\mu$.  The recent
measurement of $a_\mu$ suggests a
significant deviation from the SM prediction if the HVP contribution
is evaluated based on the $R$-ratio.  One possible
solution to such a deviation is to introduce a sizable SUSY
contribution to $a_\mu$.  However, such a possibility may conflict
with the stability of the EW vacuum because a large trilinear scalar
coupling, which is necessary to enhance the SUSY contribution to
$a_\mu$, may destabilize the EW vacuum.

We have performed a detailed analysis of the stability of the EW
vacuum in the MSSM.  We first give a complete formula to perform a
full one-loop calculation of the decay rate.  The one-loop calculation
is necessary to determine the overall factor of the decay rate as well as
to reduce its renormalization-scale dependence.  The one-loop
contribution to the decay rate can be evaluated by solving relevant
differential equations; we have given a set of differential equations for the calculation of the decay rate in the MSSM.  The counter terms to remove the
ultra-violet divergences are also given in the $\overline{\rm MS}$-scheme.

Then, we have calculated the decay rate of the EW vacuum and derived
an upper bound on the lighter smuon mass.  In our procedure, we have
numerically calculated the bounce configuration with using the
gradient-flow method.  Once we obtain the bounce, we derive the fluctuation
operators, which give the differential equations to be solved for the calculation of the
one-loop contribution.  We have numerically solved the differential
equations and calculated the one-loop effects on the decay rate with properly
removing the divergences adopting the $\overline{\rm MS}$ scheme.
The upper bound on the lighter smuon mass has been obtained as
a function of the SUSY contribution to the muon anomalous magnetic
moment.  Requiring $1\sigma$ ($2\sigma$) consistency between the
experimentally measured and theoretically predicted values of $a_\mu$
adopting the HVP contribution based on the $R$-ratio, the smuon mass is required to be lighter than $1.37\ {\rm
  TeV}$ ($1.66\ {\rm TeV}$).  Notice, however, that the upper bound
can be relaxed if we adopt the HVP contribution based on the recent lattice
results.

\vspace{2mm}
\noindent{\it Acknowledgments:} S.C. is supported by the Director,
Office of Science, Office of High Energy Physics of the
U.S. Department of Energy under the Contract No.\ DE-AC02-05CH1123.
T.M. is supported by JSPS KAKENHI Grant Number 22H01215.  Y.S. is
supported by I-CORE Program of the Israel Planning Budgeting Committee
(grant No. 1937/12).

\appendix
\section{Beta Functions}
\label{sec:beta}
\setcounter{equation}{0}

In this appendix, we summarize the one-loop beta functions of the slepton EFT used in our analysis, whose Lagrangian is given by Eq.~\eqref{eq:EFT_Lagrangian}.
Hereafter, we use a matrix notation of the EFT couplings expressed by simply suppressing the flavor indices.
Note that some of the couplings, i.e., $\lHR$, $\lHL$, $\lkappa$, $m_R^2$, and $m_L^2$, have only one flavor index and form diagonal matrices, while some of the couplings, i.e., the Yukawa couplings, $T$, $m_L^2$, and $m_R^2$, are diagonal simply because we neglect the flavor-violating effect.
(See the discussion at the end of this section for the consistency of our treatment.)
We define the beta function of a coupling constant $X$ with
\begin{align}
  \beta_X \equiv \frac{d}{dt} X,
\end{align}
where $t\equiv \ln Q$ with $Q$ being the renormalization scale.

Taking the $SU(5)$ normalization of the $U(1)_Y$ coupling, one-loop
beta functions of the gauge couplings are given by
\begin{align}
  \beta_{g_a} = \frac{b_a}{16\pi^2} g_a^3
  ~~
  (a=1,2,3),
\end{align}
with
\begin{align}
  (b_1, b_2, b_3) = \left( \frac{41}{10} + \frac{3}{20} N_f,
  -\frac{19}{6} + \frac{1}{12} N_f, -7 \right),
\end{align}
where the number of slepton flavors is $N_f=1$ $(3)$ when only smuon is considered (three generations of sleptons are considered).

One-loop beta functions of the Yukawa couplings are given by
\begin{align}
  (16\pi^2) \beta_{Y_\ell} =&
  \left(
    Y_2(H) + \frac{3}{2} \YldYl + \frac{1}{2} \YRdYR + \frac{1}{2} \YLdYL
    - \frac{9}{4} g_2^2 - \frac{9}{4} g_1^2
  \right) Y_\ell, \\
  (16\pi^2) \beta_{Y_R} =&
  \left(
    \mathrm{Tr} \left[ \YLdYL + \frac{1}{2} \YRdYR \right]
    + \YlYld + \frac{3}{2} \YRdYR
    - \frac{9}{5} g_1^2
  \right) Y_R, \\
  (16\pi^2) \beta_{Y_L} =&
  \left(
    \mathrm{Tr} \left[ \YLdYL + \frac{1}{2} \YRdYR \right]
    + \frac{1}{2} \YldYl + \frac{3}{2} \YLdYL
    - \frac{9}{4} g_2^2 - \frac{9}{20} g_1^2
  \right) Y_L,
\end{align}
with
\begin{align}
  Y_2(H) \equiv \mathrm{Tr} \left[
    3Y_u^2 + 3Y_d^2 + Y_\ell^2
  \right],
\end{align}
where $Y_u$ and $Y_d$ are the SM-like Yukawa couplings of the up-type and down-type quarks, respectively.
Note that, since we neglect the flavor- and CP-violating effects, we do not need to carefully treat the ordering of matrix products and Hermite conjugate of Yukawa coupling matrices.

One-loop beta functions of the scalar quartic couplings are given by
\begin{align}
  (16\pi^2) \beta_{\lH} =&
  24 \lH^2 + \left( 4 Y_2(H) - 9 g_2^2 - \frac{9}{5} g_1^2 \right) \lH
  + \Tr (\lHL^2 + \lHR^2 + \lkappa^2) \notag \\
  &- 2 Y_4(H) + \frac{9}{8} g_2^4 + \frac{9}{20} g_2^2 g_1^2 + \frac{27}{200} g_1^4,\\
  (16\pi^2) \beta_{\lR} =&
  4(2F - I) \circ \lR^2 + 8(F + I) \circ \lR \circ \lR + 2 \lLR \lLR^T + 2\lHR F \lHR \notag \\
  &+ 2Y_R^2 \lR + 2\lR Y_R^2 - \frac{36}{5} g_1^2 \lR
  -2 Y_R^2 F Y_R^2
  + \frac{54}{25} g_1^4 F,\\
  (16\pi^2) \beta_{\lL} =&
  4(3F - I) \circ \lL^2 + 8(F + I) \circ \lL \circ \lL + 4(\lL \lLt + \lLt \lL) \notag \\
  &+ 4\lambda_L^{'2} + 4\lLt \circ \lLt + \lLR^T \lLR + \lHL^2 + \lkappa^2+ \lHL (F-I) \lkappa + \lkappa (F-I) \lHL \notag \\
  &+ 2 Y_L^2 \lL + 2 \lL Y_L^2 - \left( 9 g_2^2 + \frac{9}{5} g_1^2 \right) \lL - 2 Y_L^2 F Y_L^2 \notag \\
  &+ \left( \frac{9}{8} g_2^4 + \frac{27}{200} g_1^4 \right) F - \frac{9}{20} g_2^2 g_1^2 (F - 2I),\\
  (16\pi^2) \beta_{\lLt} =& (F - I) \circ \Bigl(
  4(\lL \lLt + \lLt \lL) + 16\lL \circ \lLt + 8\lLt \circ \lLt - 4Y_L^4
  \Bigr) \notag \\
  &+ (\lHL - \lkappa) (F - I) (\lHL - \lkappa) \notag \\
  &+ 2 Y_L^2 \lLt + 2 \lLt Y_L^2 - \left( 9 g_2^2 + \frac{9}{5} g_1^2 \right) \lLt
  + \frac{9}{10} g_2^2 g_1^2 (F-I) \\
  (16\pi^2) \beta_{\lLR} =&
  4 \lLR \circ \lLR + 8\lLR \lL + 4I \circ \lLR \circ \lL + 4((F - I) \circ \lLR) (I \circ \lL) \notag \\
  &+ 4\lLR \lLt + 4\lR \lLR + 4I \circ \lR \circ \lLR + 4(I \circ \lR) ((F - I) \circ \lLR) \notag \\
  &+ 2\lHR F (\lHL + \lkappa) \notag \\
  &+ 2\lLR Y_L^2 + 2Y_R^2 \lLR - \left( \frac{9}{2} g_2^2 + \frac{9}{2} g_1^2\right) \lLR \notag \\
  &- 4Y_R^2 F Y_L^2
  +\frac{27}{25} g_1^4 F,
\end{align}
\begin{align}
  (16\pi^2) \beta_{\lHR} =& 4 \lHR^2 + 2f(\lLR, \lHL + \lkappa) \notag \\
  &+ f\left( 12\lH + 4\lR + 4I \circ \lR + 2Y_2(H) + 2Y_R^2 - \frac{9}{2} g_2^2 - \frac{9}{2} g_1^2, \lHR \right) \notag \\
  &- 4Y_\ell^2 Y_R^2 + \frac{27}{25} g_1^4,\\
  (16\pi^2) \beta_{\lHL} =&
  6 \lHL^2 - 4 \lHL \lkappa + 2\lkappa^2 + 4f(\lH + \lL, \lkappa) + 2f(\lLR^T, \lHR) \notag \\
  &+ f\left( 8\lH + 4\lL + 4I \circ \lL + 4\lLt + 2Y_2(H) + 2Y_L^2 - 9g_2^2 - \frac{9}{5} g_1^2, \lHL \right) \notag \\
  &+ \frac{9}{4} g_2^4 + \frac{9}{10} g_2^2 g_1^2 + \frac{27}{100} g_1^4,\\
  (16\pi^2) \beta_{\lkappa} =&
  6\lkappa^2 - 4 \lHL \lkappa + 2\lHL^2 + 4f(\lH + \lL, \lHL) + 2f(\lLR^T, \lHR) \notag \\
  &+ f\left(8\lH + 4\lL + 4I \circ \lL + 4\lLt + 2Y_2(H) + 2Y_L^2 - 9g_2^2 - \frac{9}{5} g_1^2, \lkappa \right) \notag \\
  &- 4Y_\ell^2 Y_L^2
  + \frac{9}{4} g_2^4 - \frac{9}{10} g_2^2 g_1^2 + \frac{27}{100} g_1^4,
\end{align}
where
\begin{align}
  Y_4(H) &\equiv \mathrm{Tr} \left[ 3Y_u^4 + 3Y_d^4 + Y_\ell^4 \right],
\end{align}
and $I$ and $F$ are constant matrices defined as
\begin{align}
  I \equiv
  \begin{pmatrix}
    1 & 0 & 0 \\
    0 & 1 & 0 \\
    0 & 0 & 1
  \end{pmatrix},
  ~~~
  F \equiv
  \begin{pmatrix}
    1 & 1 & 1 \\
    1 & 1 & 1 \\
    1 & 1 & 1
  \end{pmatrix}. \label{eq:IB}
\end{align}
We also use the Hadamard product defined through
\begin{align}
  (M \circ N)_{\alpha\beta} &\equiv M_{\alpha\beta} N_{\alpha\beta}, \label{eq:Hadamard}
\end{align}
and a matrix operation that maps a diagonal matrix $D$ to another diagonal matrix
\begin{align}
  f(M, D) = I \circ (M D F).
\end{align}

Finally, one-loop beta functions of dimensionful parameters are given by
\begin{align}
  (16\pi^2) \beta_{T} =& \left( Y_2(H) - \frac{9}{2} g_2^2 - \frac{27}{10} g_1^2 \right) T
  + T Y_L^2 + Y_R^2 T \notag \\
  &+ 2\lHR T + T \left( - 2\lHL + 4\lkappa \right)
  + 2\lLR \circ T - 4M_1 Y_\ell Y_R Y_L ,\\
  (16\pi^2) \beta_{M_1} =& \mathrm{Tr} \left( 2Y_L^2 + Y_R^2 \right) M_1,\\
  (16\pi^2) \beta_{m_H^2} =& \left( 12\lH + 2Y_2(H) - \frac{9}{2} g_2^2 - \frac{9}{10} g_1^2 \right) m_H^2 \notag \\
  &+ \mathrm{Tr} \left[ 2(\lHL + \lkappa) \mLSq + 2\lHR \mRSq
  + 2T^2 \right],\\
  (16\pi^2) \beta_{\mRSq} =& f\left( 4\lR + 4I \circ \lR - \frac{18}{5} g_1^2, \mRSq \right) \notag \\
  &+ 4f(\lLR, \mLSq) +  4m_H^2 \lHR
  + 2\YRYRd \mRSq - 4 \YRYRd M_1^2+ 4T^2, \label{eq:betamRSq} \\
  (16\pi^2) \beta_{\mLSq} =& f\left(8\lL + 4I \circ \lL + 4\lLt - \frac{9}{2} g_2^2 - \frac{9}{10} g_1^2, \mLSq \right) \notag \\
  &+ 2f(\lLR^T, \mRSq) + 2m_H^2 (\lHL + \lkappa) + 2\YLYLd \mLSq - 4 \YLYLd M_1^2 + 2T^2. \label{eq:betamLSq}
\end{align}
Note that, with the above beta-functions, the RG flow does not induce off-diagonal elements of the Yukawa couplings, $T$, $m_R^2$, and $m_L^2$, as far as we start from boundary conditions with vanishing off-diagonal elements.
This ensures the consistency of our treatment.

\section{Bounce}
\label{sec:bounce}
\setcounter{equation}{0}

In this appendix, we explain how we calculate the bounce configuration.  In our analysis, the bounce configuration is numerically calculated by using the gradient flow method \cite{Chigusa:2019wxb, Sato:2019axv}; in particular, we use the modified version proposed in \cite{Sato:2019axv}.

Among the various methods to obtain the bounce, the gradient flow method has an advantage in the calculation of fluctuation operators. In the case of our interest, we need to evaluate the functional determinants of the fluctuation operators of the size as large as $7\times7$.  In order to numerically solve the corresponding seven simultaneous differential equations up to a large enough $r$ (see Eqs.\ \eqref{diffeq_psi} and \eqref{diffeq_hatpsi}), the bounce configuration at a large value of $r$ should be well understood. In particular, we need to numerically follow the evolution of the functions $\psi_\ell(r)$ and $\hat\psi_\ell(r)$, whose evolution equations are given by Eqs.\ \eqref{diffeq_psi} and \eqref{diffeq_hatpsi}, respectively, until they show the same asymptotic behavior.  For the case of multi-field bounce, a precise calculation at $r\rightarrow\infty$ is highly non-trivial because, at $r\rightarrow\infty$, the functions exponentially grow with different growth rates.  The gradient flow method ensures that the solutions approach the false vacuum at $r\rightarrow\infty$ with the correct asymptotic behaviors; we found that it stabilizes the calculation of the functional determinant.

Here, the scalar fields responsible for the bounce configuration are denoted as $\rho_i$, with $i$ being the index distinguishing the field species. The field values at the true vacuum and the false vacuum are denoted by $v^{\rm (T)}_i$ and $v^{\rm (F)}_i$, respectively.

We define
\begin{align}
    \mathcal V&=2\pi^2\int_0^\infty\dd{r}r^3V(\rho(r)),\\
    \mathcal T&=2\pi^2\int_0^\infty\dd{r}r^3\sum_i\frac{1}{2}\rho'^2_i(r),
\end{align}
where $V(\rho)$ is the scalar potential with $V(v^{\rm (F)})=0$.
It has been shown that a minimization sequence of $\mathcal T$ with a fixed $\mathcal V$ converges to a non-trivial configuration, which is related to the bounce through the scale transformation \cite{Coleman:1977th}. Such a minimization sequence can be realized by the gradient flow method \cite{Sato:2019axv}. The flow equation is given by
\begin{equation}
    \partial_s\rho_i=\partial_r^2\rho_i+\frac{3}{r}\partial_r\rho_i-\lambda(\rho)\pdv{V}{\rho_i},
\end{equation}
where $s$ is the flow time and
\begin{equation}
    \lambda(\rho)=\frac{\sum_i\int_0^\infty\dd{r}r^3\pdv{V}{\rho_i}\qty[\partial_r^2\rho_i+\frac{3}{r}\partial_r\rho_i]}{\sum_i\int_0^\infty\dd{r}r^3\qty(\pdv{V}{\rho_i})^2}.
\end{equation}
Here, $\rho_i$ is promoted to a function of $r$ and the flow time $s$.

In our calculation, we use a different radius variable, which is defined as
\begin{equation}
    x=\tanh\frac{r}{r_0},
\end{equation}
where $r_0$ is a constant. This allows us to set the boundary conditions explicitly,
\begin{align}
    \partial_x\rho_i(x=0)=0,~~~\rho_i(x=1)=v^{\rm (F)}_i.
\end{align}
In terms of $x$, $\mathcal V$ is written as
\begin{equation}
    \mathcal V=2\pi^2r_0^4\int_0^1\dd{x}\frac{\arctanh^3 x}{1-x^2}V(\rho),
\end{equation}
and the flow equation becomes
\begin{equation}
    \partial_s\rho_i=\qty(1-x^2)^2\partial_x^2\rho_i+(1-x^2)\qty(\frac{3}{\arctanh x}-2x)\partial_x\rho_i-\lambda(\rho)\pdv{V}{\rho_i},\label{eq_bounce_flow}
\end{equation}
where
\begin{equation}
    \lambda(\rho)=\frac{\sum_i\int_0^1\dd{x}\pdv{V}{\rho_i}\qty[\qty(1-x^2)\partial_x^2\rho_i+\qty(\frac{3}{\arctanh x}-2x)\partial_x\rho_i]\arctanh^3 x}{\sum_i\int_0^1\dd{x}\frac{\arctanh^3 x}{1-x^2}\qty(\pdv{V}{\rho_i})^2}.
\end{equation}

We first find an initial field configuration that gives a negative value of $\mathcal V$.
We take the following configuration:
\begin{equation}
    \rho_i(x)=\tilde v_i+\frac12[\tanh \{ c_2 (\arctanh x^2 - c_3 )\}+1](v^{\rm (F)}_i-\tilde v_i),
\end{equation}
with
\begin{equation}
    \tilde v_i=c_1(v^{\rm (T)}_i-v^{\rm (F)}_i)+v^{\rm (F)}_i,
\end{equation}
where $c_i$'s are constants.  In our numerical calculation, we take $c_3=0.3$. Then, we start with $(c_1,c_2)=(0.001,3)$ and increase them until $\mathcal V<0$ is realized.

We numerically solve the flow equation until the right-hand side of Eq.~\eqref{eq_bounce_flow} becomes small enough.  The bounce solution is then obtained as 
\begin{equation}
    \rho_i(r)=\rho_i\qty(\tanh\frac{r}{\sqrt{\lambda}}).
\end{equation}
Here, $r$ is rescaled so that the solution satisfies the Derrick's relation, $\mathcal T=-2\mathcal V$.  Notice that all the above formulas are independent of $r_0$ and thus we do not need to know the typical size of the bounce before the calculation.

With the bounce configuration given above, the Euclidean action is given by
\begin{align}
    S_E&=2\pi^2\lambda^2\int_0^1\dd{x}\frac{\arctanh^3x}{1-x^2}\qty[\frac{(1-x^2)^2}{2\lambda}(\partial_x\rho)^2+V(\rho(x))].
\end{align}

\section{Fluctuation Operators}
\label{sec:fluc_op}
\setcounter{equation}{0}

In this appendix, we summarize the fluctuation operators that we use in our calculation.
We decompose all the scalar fields into canonically normalized real scalar fields, denoted as 
$\Phi_i$ (with $i$ being the index distinguishing the species of the 
scalars).  The scalars are expanded around the bounce, 
$\rho_i (r)$, as
\begin{equation}
  \Phi_i=\rho_i+\varphi_i,
\end{equation}
where $\varphi_i$ is the fluctuation around the bounce.

The fluctuation operators for sleptons with $\rho_i\neq0$ are different from those with $\rho_i=0$. In the following, we use index $\rho$ for the sleptons with $\rho_i\neq0$ and $I$ for those with $\rho_i=0$.
More explicitly, when we consider the instability towards the smuon direction assuming the other sfermions are heavy, $\rho=2$ and there is no contribution from other sfermions (i.e., the index $I$ is irrelevant). When we consider the instability towards the stau direction including the other sleptons, $\rho=3$ and $I=1,2$.

The fluctuation operators, $\mathcal M^{(A_\mu\varphi)}$, $\mathcal M^{(c\bar c)}$ and $\mathcal M^{(\psi)}$, are large matrices containing all the fields in EFT, where 
$\mathcal M^{(A_\mu\varphi)}$, $\mathcal M^{(c\bar c)}$ and $\mathcal M^{(\psi)}$ are those of 
the gauge bosons and the scalars, the ghosts, and fermions, respectively. 
However, in practice, they can be block-diagonalized and we can calculate them one by one. We give each block below.

\subsection{Scalars that do not mix with gauge bosons}

Concerning $\mathcal M^{(A_\mu\varphi)}$, we find  blocks that contain only scalar fluctuations. (Such scalars are denoted by $\phi_i$.)
Although the fluctuation operators for them can be derived from the formulas with gauge bosons given in the next subsection, we discuss them separately since they have a much simpler structure.

The bounce-dependent mass terms in each block can be denoted as
\begin{equation}
    \mathcal L_\phi = 
    -\frac12\Omega_{ij}^{(\phi)}(\rho)\phi_i\phi_j.
\end{equation}
Then, the fluctuation operator is given by
\begin{equation}
    \mathcal M^{(\phi)}=-\partial^2+\Omega^{(\phi)}.
\end{equation}
We expand the scalar fields by using mode functions as
\begin{equation}
    \phi=\alpha_{\ell m_1m_2}^\phi(r)Y_{\ell m_1m_2},
\end{equation}
where the sum over $\ell,m_1$ and $m_2$ are implicit.  Here, $Y_{\ell
  m_1m_2}$ is the hyperspherical harmonics where $\ell$, which is an integer, is the 4D angular momentum quantum number and $m_1$ and $m_2$ are the
quantum numbers that correspond to the magnetic quantum number.  Then, 
\begin{equation}
  \partial^2 \alpha_{\ell m_1m_2}^\phi Y_{\ell m_1m_2}
  =\qty(\Delta_\ell \alpha_{\ell m_1m_2}^\phi) Y_{\ell m_1m_2},
\end{equation}
where
\begin{equation}
    \Delta_\ell \equiv \partial_r^2+\frac{3}{r}\partial_r-\frac{\ell(\ell+2)}{r^2}.
\end{equation}
Since the bounce is $O(4)$-symmetric, there is no mixing among different angular momenta and the contribution to the prefactor is given by
\begin{equation}
    \mathcal A^{(\phi)}=-\frac12\ln\frac{\det\mathcal M^{(\phi)}}{\det\widehat{\mathcal M}^{(\phi)}}=-\frac12\sum_{\ell=0}^\infty(\ell+1)^2\ln\frac{\det\mathcal M_\ell^{(\phi)}}{\det\widehat{\mathcal M}_\ell^{(\phi)}},
\end{equation}
where
\begin{equation}
    \mathcal M_\ell^{(\phi)}=-\Delta_\ell+\Omega^{(\phi)}.
\end{equation}
Notice that $(\ell+1)^2$ is the degeneracy factor due to $m_1$ and $m_2$.

\subsubsection{Scalars with $\rho'_i(r)\neq0$}

We first consider fluctuations of the fields which the bounce consists of, i.e., $\Re H^0$, $\Re \tilde \ell_{L\rho}^-$, and $\Re \tilde \ell_{R\rho}$. Because of the $U(1)_{\rm EM}$ breaking due to the bounce, neutral and charged components mix with each other.  They include the fluctuations corresponding to the translation of the bounce, and hence there appear translational zero modes for $\ell=1$. 

In the basis of $(\Re H^0,\Re \tilde \ell_{L\rho}^-,\Re \tilde \ell_{R\rho})/\sqrt{2}$, the bounce-dependent mass matrix is given by
\begin{align}
    \Omega_{11}^{(\rho)}&=m_H^2+\frac{\lambda_{HR\rho}}{2}\rho_R^2+3\lambda_H\rho_h^2+\frac{\kappa'_{\rho}}{2}\rho_L^2,\\
    \Omega_{12}^{(\rho)}&=-\frac{T_\rho}{\sqrt{2}}\rho_e+\kappa'_{\rho}\rho_h\rho_L,\\
    \Omega_{13}^{(\rho)}&=-\frac{T_\rho}{\sqrt{2}}\rho_L+\lambda_{HR\rho}\rho_R\rho_h,\\
    \Omega_{22}^{(\rho)}&=m_{L\rho}^2+\frac{\lambda_{LR\rho\rho}}{2}\rho_R^2+\frac{\kappa'_{\rho}}{2}\rho_h^2+3\lambda_{L\rho\rho}\rho_L^2,\\
    \Omega_{23}^{(\rho)}&=-\frac{T_\rho}{\sqrt{2}}\rho_h+\lambda_{LR\rho\rho}\rho_R\rho_L,\\
    \Omega_{33}^{(\rho)}&=m_{R\rho}^2+3\lambda_{R\rho\rho}\rho_R^2+\frac{\lambda_{HR\rho}}{2}\rho_h^2+\frac{\lambda_{LR\rho\rho}}{2}\rho_L^2.
\end{align}

\subsubsection{Sleptons with $\rho_i(r)=0$}

In the case of three generations of sleptons, $\tilde\ell_{LI}$ and $\tilde\ell_{RI}$ do not mix with gauge bosons.  Fluctuation operators of these fields can be decomposed into those for the neutral and charged fields.

The neutral fields, $\Re\tilde \ell_{LI}^0/\sqrt{2}$ and $\Im\tilde \ell_{LI}^0/\sqrt{2}$, have the bounce-dependent mass of
\begin{equation}
    \Omega^{(N)}=
    m_{LI}^2+\frac{\lambda_{LR\rho I}}{2}\rho_R^2+\frac{\lambda_{HLI}}{2}\rho_h^2+\lambda_{L\rho I}\rho_L^2.
\end{equation}

The fluctuation operators of the charged fields become $2\times 2$; those of $(\Re\tilde\ell_{LI}^-,\Re\tilde\ell_{RI})/\sqrt{2}$ and $(-\Im\tilde
\ell_{LI}^-,\Im\tilde\ell_{RI})/\sqrt{2}$ are both derived by using the following bounce-dependent mass matrix:
\begin{equation}
    \Omega^{(C)}=
    \mqty(m_{LI}^2+\frac{\lambda_{LR\rho I}}{2}\rho_R^2+\frac{\kappa'_{I}}{2}\rho_h^2+(\lambda_{L\rho I}+\lambda'_{L\rho I})\rho_L^2
        &-\frac{T_I}{\sqrt{2}}\rho_h\\
        -\frac{T_I}{\sqrt{2}}\rho_h&m_{RI}^2+\lambda_{R\rho I}\rho_R^2+\frac{\lambda_{HRI}}{2}\rho_h^2+\frac{\lambda_{LRI\rho}}{2}\rho_L^2
    ).
\end{equation}
\subsection{Gauge bosons and scalars that mix with each other}
Here, we give the fluctuation operators for the blocks that have both the gauge boson fluctuations and the scalar fluctuations.
We work in the background gauge with the gauge fixing parameter being $\xi=1$. The gauge fixing terms are given by
\begin{equation}
    \mathcal L_{\rm GF}=\frac{1}{2}\sum_a(\partial^\mu A_\mu^a-\varphi^iM_{ia})^2,
\end{equation}
where 
\begin{equation}
    M_{ia}=-g_a\tau^a_{ij}\rho_j,
\end{equation}
and $\tau^a_{ij}$ is the generators of the gauge group acting on real
fields:
\begin{equation}
  D_\mu\Phi_i \equiv \partial_\mu\Phi_i+\sum_{a} g_aA^a_\mu \tau^a_{ij}\Phi_j.
\end{equation}
Here, $g_a$'s are the gauge couplings and $A^a_\mu$'s are the gauge
bosons, namely the photon, the $Z$ boson and the $W$ bosons.  The index
$a$ runs over all the gauge bosons and $i$ runs over all the
scalars. Since we do not break the $SU(3)_C$ symmetry, the gluons do
not contribute to the decay rate at the one-loop level.  In addition, the bounce-dependent
mass terms of the Nambu-Goldstone bosons (and the scalars that mix
with them) are given in the following form:
\begin{equation}
  \mathcal L = 
  -\frac12\Omega_{ij}^{\rm (NG)}(\rho)\varphi_i\varphi_j.
\end{equation}
Then, the fluctuation operator of scalar and gauge bosons are given by
\begin{equation}
    \mathcal M^{(A_\mu\varphi)}=\mqty((-\partial^2+M^tM)\delta_{\mu\nu}&2(\partial_\nu M)^t\\
    2(\partial_\nu M)&-\partial^2+\Omega^{\rm (NG)}+MM^t).
\end{equation}

There also exist contributions from the Faddeev-Popov ghosts. The
fluctuation operator of the ghosts is given by
\begin{equation}
    \mathcal M^{(c\bar c)}=-\partial^2+M^tM,
\end{equation}
which has the degeneracy of two. 

We expand the scalar fields and the Faddeev-Popov ghosts with the
hyperspherical functions, while the gauge bosons are expanded as
\begin{align}
    A^a_\mu=&\,\alpha^a_{S\ell m_1m_2}(r)\frac{x_\mu}{r}Y_{\ell m_1m_2}+\alpha^a_{L\ell m_1m_2}(r)\frac{r}{L}\partial_\mu Y_{\ell m_1m_2}\nonumber\\
    &+\alpha^a_{T1\ell m_1m_2}(r)i\epsilon_{\mu\nu\rho\sigma}V_\nu^{(1)}L_{\rho\sigma} Y_{\ell m_1m_2}+\alpha^a_{T2\ell m_1m_2}(r)i\epsilon_{\mu\nu\rho\sigma}V_\nu^{(2)}L_{\rho\sigma} Y_{\ell m_1m_2},
\end{align}
where $V_\nu^{(i)}$'s are arbitrary independent vectors and
\begin{equation}
    L_{\mu\nu}=\frac{i}{\sqrt{2}}(x_\mu\partial_\nu-x_\nu\partial_\mu).
\end{equation}
The $\ell>0$ contributions from the ghost and the contributions from $\alpha^a_{T1\ell m_1m_2}$ and $\alpha^a_{T2\ell m_1m_2}$ are canceled out and we are left with
\begin{align}
    \mathcal A^{(A_\mu\varphi c\bar c)}&=\ln\frac{\det\mathcal M^{(c\bar c)}}{\det\widehat{\mathcal M}^{(c\bar c)}}-\frac12\ln\frac{\det\mathcal M^{(A_\mu\varphi)}}{\det\widehat{\mathcal M}^{(A_\mu\varphi)}}\nonumber\\
    &=\ln\frac{\det\mathcal M_0^{(c\bar c)}}{\det\widehat{\mathcal M}_0^{(c\bar c)}}-\frac12\ln\frac{\det\mathcal M_0^{(S\varphi)}}{\det\widehat{\mathcal M}_0^{(S\varphi)}}-\frac12\sum_{\ell=1}^\infty(\ell+1)^2\ln\frac{\det\mathcal M_\ell^{(SL\varphi)}}{\det\widehat{\mathcal M}_\ell^{(SL\varphi)}},
\end{align}
where
\begin{align}
    \mathcal M_0^{(c\bar c)} =&\, -\Delta_0+M^TM,
    \\
    \mathcal M_0^{(S\varphi)}=&\, \mqty(-\Delta_1+M^TM&2(M')^T\\2M'&-\Delta_0+\Omega^{({\rm NG})}+MM^T),
    \\
    \mathcal M_\ell^{(SL\varphi)}=&\, \mqty(-\Delta_{\ell-1}+M^TM&0&\sqrt{\frac{2\ell}{\ell+1}}(M')^T\\0&-\Delta_{\ell+1}+M^TM&-\sqrt{\frac{2(\ell+2)}{\ell+1}}(M')^T\\\sqrt{\frac{2\ell}{\ell+1}}M'&-\sqrt{\frac{2(\ell+2)}{\ell+1}}M'&-\Delta_\ell+\Omega^{({\rm NG})}+MM^T).
\end{align}
We block-diagonalize the above operators; each block is given below.

\subsubsection{Charged gauge bosons}

Here, we give the fluctuation operators for $W^{1}$, $W^{2}$,
and the scalars that mix with them.
For $W^1$, the basis of the scalars is $(\Im H^+,\Im \ell_{L\rho}^0)/\sqrt{2}$.
For $W^2$, it is $(\Re H^+,\Re \tilde \ell_{L\rho}^0)/\sqrt{2}$. The fluctuation operator is constructed by
\begin{equation}
    M^{(W\pm)}=\mqty(\frac{g_2}{2}\rho_h\\\frac{g_2}{2}\rho_L),
\end{equation}
and
\begin{equation}
    \Omega^{(W\pm)}=\mqty(
        m_H^2+\frac{\lambda_{HR\rho}}{2}\rho_R^2+\lambda_H\rho_h^2+\frac{\lambda_{HL\rho}}{2}\rho_L^2&
        -\frac{T_\rho}{\sqrt{2}}\rho_R+\frac{\kappa_{\rho}}{2}\rho_h\rho_L\\
        -\frac{T_\rho}{\sqrt{2}}\rho_R+\frac{\kappa_{\rho}}{2}\rho_h\rho_L&
        m_{L\rho}^2+\frac{\lambda_{HL\rho}}{2}\rho_h^2+\frac{\lambda_{LR\rho\rho}}{2}\rho_R^2+\lambda_{L\rho\rho}\rho_L^2
    ).
\end{equation}
\subsubsection{Neutral gauge bosons}
The two neutral gauge bosons become massive and mix with each other
because of the $U(1)_{\rm EM}$ symmetry breaking due to the bounce. We
take the basis of $(W^3,B)$ for the gauge bosons, and $(\Im H^0,\Im
\tilde \ell_{L\rho}^-,\Im\tilde\ell_{R\rho})/\sqrt{2}$ for the
scalars. The fluctuation operator is constructed by
\begin{equation}
    M^{(Z\gamma)}=\mqty(
        -\frac{g_2}{2}\rho_h
        &\sqrt{\frac{3}{5}}\frac{g_1}{2}\rho_h\\
        -\frac{g_2}{2}\rho_L
        &-\sqrt{\frac{3}{5}}\frac{g_1}{2}\rho_L\\
        0
        &\sqrt{\frac{3}{5}}g_1\rho_R
    ),\label{eq_apx_neutral_m}
\end{equation}
and
\begin{align}
    \Omega_{11}^{(Z\gamma)}&=m_H^2+\frac{\lambda_{HR\rho}}{2}\rho_R^2+\lambda_H\rho_h^2+\frac{\kappa'_{\rho}}{2}\rho_L^2,\\
    \Omega_{12}^{(Z\gamma)}&=-\frac{T_\rho}{\sqrt{2}}\rho_e,\\
    \Omega_{13}^{(Z\gamma)}&=-\frac{T_\rho}{\sqrt{2}}\rho_L,\\
    \Omega_{22}^{(Z\gamma)}&=m_{L\rho}^2+\frac{\lambda_{LR\rho\rho}}{2}\rho_R^2+\frac{\kappa'_{\rho}}{2}\rho_h^2+\lambda_{L\rho\rho}\rho_L^2,
    \\
    \Omega_{23}^{(Z\gamma)}&=\frac{T_\rho}{\sqrt{2}}\rho_h,\\
    \Omega_{33}^{(Z\gamma)}&=m_{R\rho}^2+\lambda_{R\rho\rho}\rho_R^2+\frac{\lambda_{HR\rho}}{2}\rho_h^2+\frac{\lambda_{LR\rho\rho}}{2}\rho_L^2.
\end{align}

There appears a zero mode in $\ell=0$ in association with the $U(1)_{\rm EM}$ breaking.
Following \cite{Chigusa:2020jbn}, the Jacobian for the gauge zero mode is given by
\begin{equation}
    \mathcal J_{\rm EM}=\qty(\frac{e^2}{\pi\det\mathcal K})^{-1/2},
\end{equation}
where
\begin{equation}
    \mathcal K=\lim_{r\to\infty} r^3\mathcal U^T(\partial_r\psi^{(c\bar c)}_0)(\psi^{(c\bar c)}_0)^{-1}\mathcal U.
\end{equation}
Here, $\psi^{(c\bar c)}_0$ is the solution of $\mathcal M_0^{(c\bar
  c)}\psi^{(c\bar c)}_0=0$. In addition, $\mathcal U$ is a vector
satisfying $\widehat M^{(Z\gamma)}\mathcal U=0$ and $|\mathcal U|=1$,
where $\widehat M^{(Z\gamma)}$ is given by
Eq.~\eqref{eq_apx_neutral_m} evaluated at $r\rightarrow\infty$ (i.e.,
the false vacuum).

\subsection{Fermions}

We also consider the fluctuation operators of the fermions with 
block-diagonalizing the bounce-dependent fermion mass matrix. For each block, the mass terms can be expressed as
\begin{equation}
    \mathcal L=-\frac12m_{ij}^{(\psi)}(\rho)\bar\psi_i\psi_j,
\end{equation}
where $\psi_i$'s are the 4-component Weyl fermion and $m^{(\psi)}$ is a real-valued symmetric matrix.
Then, the fluctuation operator is given by
\begin{align}
    \mathcal M^{(\psi)}&=\cancel \partial+m^{(\psi)}.
\end{align}
We do not consider the mass terms that are proportional to $\gamma^5$ since they do not exist in the present setup.

We can expand the fermionic fluctuations with the eigenfunctions characterized by $(K,K'=K+1/2,m_K,m_{K'})$ and $(K,K'=K-1/2,m_K,m_{K'})$, where $K$ and $K'$ are the total spin quantum numbers for $so(4)\simeq su(2)\times su(2)$, and $m_K$ and $m_{K'}$ are the second spin quantum numbers for them
\cite{Avan:1985eg}. Notice that $K$ and $K'$ differ by $1/2$ to construct the spinor representation. We have two independent eigenvectors for each: $\Psi_{L\pm\frac12,L,m_K,m_L,\lambda,i}$ with $\lambda=1,2$.
The fluctuation operator acts on these states as
\begin{equation}
    \mathcal M^{(\psi)}
    \begin{pmatrix}
        \Psi_{K,K+\frac12,m_K,m_L,1}\\
        \Psi_{K,K+\frac12,m_K,m_L,2}
    \end{pmatrix}=
    \begin{pmatrix}
        \partial_r-\frac{2K}{r}&m^{(\psi)}\\
        m^{(\psi)}&\partial_r+\frac{2K+3}{r}
    \end{pmatrix}
    \begin{pmatrix}
        \Psi_{K,K+\frac12,L,m_K,m_L,1}\\
        \Psi_{K,K+\frac12,L,m_K,m_L,2}
    \end{pmatrix},
\end{equation}
for $K'=K+1/2$ and 
\begin{equation}
    \mathcal M^{(\psi)}
    \begin{pmatrix}
        \Psi_{K,K-\frac12,m_K,m_L,1}\\
        \Psi_{K,K-\frac12,m_K,m_L,2}
    \end{pmatrix}=
    \begin{pmatrix}
        \partial_r+\frac{2K+2}{r}&m^{(\psi)}\\
        m^{(\psi)}&\partial_r-\frac{2K-1}{r}
    \end{pmatrix}
    \begin{pmatrix}
        \Psi_{K,K-\frac12,m_K,m_L,1}\\
        \Psi_{K,K-\frac12,m_K,m_L,2}
    \end{pmatrix},
\end{equation}
for $K'=K-1/2$.
Then, the second derivative operators are obtained as
\begin{align}
    \qty[\det\mathcal M^{(\psi)}_{K,K+\frac12,m_K,m_L}]^2&=\det\qty[\begin{pmatrix}
        \partial_r-\frac{2K}{r}&m^{(\psi)}\\
        m^{(\psi)}&\partial_r+\frac{2K+3}{r}
    \end{pmatrix}
    \begin{pmatrix}
        -\partial_r-\frac{2K+3}{r}&m^{(\psi)}\\
        m^{(\psi)}&-\partial_r+\frac{2K}{r}
    \end{pmatrix}]\nonumber\\
    &=\det
    \begin{pmatrix}
        -\Delta_{2K+1}+m^{(\psi)}m^{(\psi)}&\partial_rm^{(\psi)}\\
        \partial_rm^{(\psi)}&-\Delta_{2K}+m^{(\psi)}m^{(\psi)}
    \end{pmatrix},
\end{align}
and
\begin{align}
    \qty[\det\mathcal M^{(\psi)}_{K,K-\frac12,m_K,m_L}]^2&=\det\qty[\begin{pmatrix}
        \partial_r+\frac{2K+2}{r}&m^{(\psi)}\\
        m^{(\psi)}&\partial_r-\frac{2K-1}{r}
    \end{pmatrix}
    \begin{pmatrix}
        -\partial_r+\frac{2K-1}{r}&m^{(\psi)}\\
        m^{(\psi)}&-\partial_r-\frac{2K+2}{r}
    \end{pmatrix}]
    \nonumber\\
    &=\det
    \begin{pmatrix}
        -\Delta_{2K-1}+m^{(\psi)}m^{(\psi)}&\partial_rm^{(\psi)}\\
        \partial_rm^{(\psi)}&-\Delta_{2K}+m^{(\psi)}m^{(\psi)}
    \end{pmatrix}.
\end{align}
Here, $m^{(\psi)}m^{(\psi)}$ indicates the matrix multiplication. Notice that the determinant of the fluctuation operator is invariant under $\cancel\partial\to-\cancel\partial$.
Since these two determinants are symmetric under $K\leftrightarrow K'$, we combine these two and obtain
\begin{equation}
    \ln\mathcal A^{(\psi)}=\sum_{\ell=0}^\infty\frac{(\ell+1)(\ell+2)}{2}\ln\frac{\det\mathcal M_\ell^{(\psi)}}{\det\mathcal {\widehat M}_\ell^{(\psi)}},
\end{equation}
where
\begin{equation}
    \mathcal M_\ell^{(\psi)}=\mqty(-\Delta_\ell+m^{(\psi)}m^{(\psi)}&\partial_rm^{(\psi)}\\\partial_rm^{(\psi)}&-\Delta_{\ell+1}+m^{(\psi)}m^{(\psi)}).
\end{equation}
The mass matrix for each block is given below.

\subsubsection{Top quark}

The top quark couples to the Higgs boson and hence contributes to the prefactor.
In the basis of $(t_L,t_R)$, the bounce-dependent mass matrix is given by
\begin{equation}
    m^{(t)}=\mqty(0&\frac{y_t}{\sqrt{2}}\rho_h\\\frac{y_t}{\sqrt{2}}\rho_h&0).
\end{equation}
Notice that there exist three copies due to the color charge.
\subsubsection{Bino and leptons}
The Bino and leptons mix with each other due to the charge-breaking bounce.
In the basis of $(\tilde B,\ell_{L\rho},\ell_{R\rho})$, the mass matrix is given by
\begin{equation}
    m^{(\tilde B\ell)}=\mqty(
    M_1&-\frac{y_{L\rho}}{\sqrt{2}}\rho_L&-\frac{y_{R\rho}}{\sqrt{2}}\rho_R\\
    -\frac{y_{L\rho}}{\sqrt{2}}\rho_L&0&0\\
    -\frac{y_{R\rho}}{\sqrt{2}}\rho_R&0&0).
\end{equation}

\section{Counter Terms}
\label{sec:ct}
\setcounter{equation}{0}

In this appendix, we evaluate the divergent part, $s_{\rm \overline{MS}}$, which is introduced in Eq.~\eqref{eq_prefactor_renormalization}.

We consider general fluctuation operators, $\mathcal M$ and $\widehat{\mathcal M}$, which are those around the bounce and the false vacuum, respectively. We define $\delta\mathcal M$ as
\begin{equation}
\delta\mathcal M \equiv \mathcal M - \widehat{\mathcal M}.
\end{equation}
Then, we obtain
\begin{equation}
  \ln\frac{\det\mathcal M}{\det\widehat{\mathcal M}}=\tr\widehat{\mathcal M}^{-1}\delta\mathcal M-\frac12\tr\widehat{\mathcal M}^{-1}\delta\mathcal M\widehat{\mathcal M}^{-1}\delta\mathcal M+\frac13\tr\widehat{\mathcal M}^{-1}\delta\mathcal M\widehat{\mathcal M}^{-1}\delta\mathcal M\widehat{\mathcal M}^{-1}\delta\mathcal M+\cdots.
\end{equation}
On the right-hand side of the above expression, only the first two terms diverge. Comparing them with Eq.~\eqref{eq_prefactor_renormalization}, we can find
\begin{align}
    s \equiv \sum_\ell d_\ell s_\ell=\tr\widehat{\mathcal M}^{-1}\delta\mathcal M-\frac12\tr\widehat{\mathcal M}^{-1}\delta\mathcal M\widehat{\mathcal M}^{-1}\delta\mathcal M.
\end{align}
The trace can be evaluated by performing the momentum integration.  Then, subtracting the divergences adopting the $\overline{\rm MS}$ scheme, $s_{\overline{\rm MS}}$ is obtained. Contributions of the fields of our interest are given below.

\subsection{Scalars that do not mix with gauge bosons}

We first consider scalar fields which do not mix with gauge bosons.
We rewrite the $\Omega^{(\phi)}$ matrix introduced in Appendix \ref{sec:fluc_op}
as
\begin{align}
    \Omega^{(\phi)}&=\widehat \Omega^{(\phi)}+\delta\Omega^{(\phi)}(r).
\end{align}
Here, we take the basis in which $\widehat \Omega^{(\phi)}$ is
diagonal.  Then, denoting the Fourier transformation of $\delta\Omega^{(\phi)}$
as $\widetilde{\delta \Omega}^{(\phi)}$, we obtain
\begin{align}
    s^{(\phi)}&=\sum_i\widetilde{\delta \Omega}^{(\phi)}_{ii}(0)I_1\qty(\widehat\Omega^{(\phi)}_{ii})
    -\frac12\sum_{ij}I_2\qty(\widehat\Omega^{(\phi)}_{ii},\widehat\Omega^{(\phi)}_{jj},\qty[\widetilde{\delta \Omega}^{(\phi)}_{ij}(k-p)]^2).
\end{align}
Using the dimensional regularization with
the space-time dimension $D=4-2\epsilon$,
the divergent integrals are given by
\begin{align}
    I_1(m)=\int\frac{\dd[D]{k}}{(2\pi)^D}\frac{1}{k^2+m^2}=-\frac{m^2}{16\pi^2}\qty(\frac{1}{\bar\varepsilon}+1-\ln\frac{m^2}{Q^2}),
\end{align}
and
\begin{align}
    &I_2(m,M,F)=\int\frac{\dd[D]{k}}{(2\pi)^4}\int\frac{\dd[D]{p}}{(2\pi)^4}\frac{1}{k^2+m^2}\frac{1}{p^2+M^2}F(|k-p|)\nonumber\\
    &=\frac{1}{128\pi^4}\int_0^\infty\dd{k}k^3F(k)\qty(\frac{1}{\bar\varepsilon}+2-\frac12\ln\frac{m^2M^2}{Q^4}+\frac{m^2-M^2}{2k^2}\ln\frac{m^2}{M^2}-\frac{\omega^2}{2k^2}\ln\frac{G_+}{G_-}),
\end{align}
where
\begin{align}
    \omega^2&=\sqrt{k^4+2k^2(m^2+M^2)+(m^2-M^2)^2},\\
    G_\pm&=k^2+m^2+M^2\pm\omega^2.
\end{align}
Here, the renormalization scale is denoted by $Q$. Subtracting terms proportional to $\bar{\varepsilon}^{-1}$ from the above expressions, $s^{(\phi)}_{\rm \overline{MS}}$ is obtained.

\subsection{Gauge bosons and scalars that mix with each other}

Next, we consider the gauge bosons and the scalars that mix with each other.  We rewrite the $M$ and $\Omega^{\rm(NG)}$ matrices as
\begin{align}
    M&=\widehat M+\delta M(r),\\
    \Omega^{\rm(NG)}&=\widehat \Omega^{\rm(NG)}+\delta\Omega^{\rm(NG)}(r),
\end{align}
where the hat indicates the matrix at the false vacuum.
We also define
\begin{align}
    \widetilde{\delta M_g^2}(k)&=\int\dd[4]{x}e^{-ikx}(\delta M^T\widehat M+\widehat M^T\delta M+\delta M^T\delta M),\\
    \widetilde{\delta M_s^2}(k)&=\int\dd[4]{x}e^{-ikx}(\delta \Omega^{\rm(NG)}+\delta M\widehat M^T+\widehat M\delta M^T+\delta M\delta M^T),\\
    \widetilde{\delta M}(k)&=\int\dd[4]{x}e^{-ikx}\delta M.
\end{align}
Here, we take the field basis so that $\widehat M^T\widehat M$ and $\widehat\Omega+\widehat M\widehat M^T$ are diagonal. Then, we obtain
\begin{align}
  s^{(A\varphi)}= &\,\delta^\mu_\mu\sum_a\widetilde{\delta M_g^2}_{aa}(0)I_1\qty(\qty[\widehat M^T\widehat M]_{aa})
\nonumber\\&
-\frac{\delta^\mu_\mu}{2}\sum_{ab}I_2\qty(\qty[\widehat M^T\widehat M]_{aa},\qty[\widehat M^T\widehat M]_{bb},\qty[\widetilde{\delta M_g^2}_{ab}(k-p)]^2)\nonumber\\
  &+\sum_i\widetilde{\delta M_s^2}_{ii}(0)I_1\qty(\qty[\widehat\Omega^{\rm(NG)}+\widehat M\widehat M^T]_{ii})\nonumber\\
  &-\frac12\sum_{ij}I_2\qty(\qty[\widehat\Omega^{\rm(NG)}+\widehat M\widehat M^T]_{ii},\qty[\widehat\Omega^{\rm(NG)}+\widehat M\widehat M^T]_{jj},\qty[\widetilde{\delta M_s^2}_{ij}(k-p)]^2)\nonumber\\
  &-4\sum_{ia}I_2\qty(\qty[\widehat M^T\widehat M]_{aa},\qty[\widehat\Omega^{\rm(NG)}+\widehat M\widehat M^T]_{ii},\qty[\widetilde{\delta M}_{ia}(k-p)]^2),
\end{align}
and
\begin{align}
    s^{(c\bar c)}&=\sum_a\widetilde{\delta M_g^2}_{aa}(0)I_1\qty(\qty[\widehat M^T\widehat M]_{aa})-\frac12\sum_{ab}I_2\qty(\qty[\widehat M^T\widehat M]_{aa},\qty[\widehat M^T\widehat M]_{bb},\qty[\widetilde{\delta M_g^2}_{ab}(k-p)]^2),
\end{align}
where $\delta^\mu_\mu=D=4-2\varepsilon$ is used.

\subsection{Fermions}

Finally, we consider the contributions of fermions. We expand the
$m^{(\psi)}$ as
\begin{equation}
  m^{(\psi)} = \widehat m^{(\psi)} +\delta m^{(\psi)} (r),
\end{equation}
where $\widehat m^{(\psi)}$ is the fermion mass matrix around the
false vacuum.  We work in the basis in which $\widehat m^{(\psi)}$ is diagonal.
Then, we obtain
\begin{align}
  s^{(\psi)} = &\, 
  2\sum_i\widetilde {\delta M_\psi}_{ii}^2(0)I_1
  \qty( \qty(\widehat m_{ii}^{(\psi)})^2 )
  \nonumber \\ &\,
  -\sum_{ij}I_2\qty( \qty(\widehat m_{ii}^{(\psi)})^2,
  \qty(\widehat m_{jj}^{(\psi)})^2,
  \qty[\widetilde {\delta M}_{\psi ij}^{2}(k-p)]^2+\qty[\widetilde{\delta m}_{ij}^{(\psi)}(k-p)]^2(k-p)^2),
\end{align}
where 
\begin{align}
    \widetilde {\delta M}_\psi^{2}(k)&=\int \dd[4]{x}e^{-ikx}
    \left[
      2\widehat m^{(\psi)} \delta m^{(\psi)}+\qty(\delta m^{(\psi)})^2
    \right].
\end{align}


\bibliographystyle{jhep}
\bibliography{ewvacmumdm}


\end{document}